\documentclass[manuscript,screen]{acmart}

\usepackage{algorithmic}
\usepackage{graphicx}
\usepackage{textcomp}
\usepackage{xcolor}
\usepackage{comment}
\usepackage{hyperref}
\usepackage{bm}
\usepackage{enumitem}
\usepackage{appendix}

\usepackage{amsmath}
\usepackage{lipsum}
\usepackage{multirow}
\usepackage[justification=centering]{caption}
\usepackage{caption}
\usepackage{url}
\usepackage{listings}
\usepackage{subfigure}
\usepackage{array}
\usepackage{pifont}
\usepackage{tikz}
\usepackage{pgf-pie}
\usepackage{pgfplots} 
\usepackage{subcaption}
\usepackage[table]{xcolor} 
\usepackage{longtable}     
\usepackage{wrapfig}
\usepackage{tikz}
\usepackage{pgf-pie}
\usepackage{pgfplots}
\usepackage[utf8]{inputenc}
\usepackage{makecell}
\usepackage{pgfplotstable}
\pgfplotsset{compat=1.18}

\usepackage{soul}

\usepackage[framemethod=tikz]{mdframed}
\mdfdefinestyle{customstyle}{
  linecolor=gray!100,
  linewidth=3pt,
  innerleftmargin=3pt,
  topline=false,
  rightline=false,
  bottomline=false,
  leftline=true,
  innerrightmargin=3pt,
  innertopmargin=3pt,
  innerbottommargin=3pt,
  backgroundcolor=gray!15
}

\newenvironment{custommdframed}
  {\begin{mdframed}[style=customstyle]}
  {\end{mdframed}}

\begin{document}

\title[An Empirical Study of Industrial Needs and Academic Capabilities in AI-Driven Software Engineering]{Aligning Academia with Industry: An Empirical Study of Industrial Needs and Academic Capabilities in AI-Driven Software Engineering}

\author{Hang Yu}
\affiliation{
  \institution{Beihang University}
   \department{SKLCCSE}
  \city{Beijing}
  \country{China}
}
\email{sy2406231@buaa.edu.cn}

\author{YuZhou Lai}
\affiliation{
  \institution{Beihang University}
   \department{SKLCCSE}
  \city{Beijing}
  \country{China}
}
\email{laiyuzhou@buaa.edu.cn}

\author{Li Zhang}
\affiliation{
  \institution{Beihang University}
   \department{SKLCCSE}
  \city{Beijing}
  \country{China}
}
\email{lily@buaa.edu.cn}

\author{Xiaoli Lian}
\authornote{This author is the corresponding author.}
\affiliation{
  \institution{Beihang University}
  \department{SKLCCSE}
  \city{Beijing}
  \country{China}
}
\email{lianxiaoli@buaa.edu.cn}

\author{Fang Liu}
\affiliation{
  \institution{Beihang University}
  \city{Beijing}
  \country{China}
}
\email{fangliu@buaa.edu.cn}

\author{Yanrui Dong}
\affiliation{
  \institution{Beihang University}
   \department{School of Computer Science and Engineering}
  \city{Beijing}
  \country{China}
}
\email{dyr24371159@buaa.edu.cn}

\author{Ting Zhang}
\affiliation{
  \institution{Monash University}
  \country{Australia}
}
\email{ting.zhang@monash.edu}

\author{Zhi Jin}
\affiliation{
  \institution{Peking University}
  \city{Beijing}
  \country{China}
}
\email{zhijin@pku.edu.cn}

\author{David Lo}
\affiliation{
  \institution{Singapore Management University}
  \country{Singapore}
}
\email{davidlo@smu.edu.sg}

\begin{abstract}

The rapid advancement of large language models (LLMs) is fundamentally reshaping software engineering (SE), driving a paradigm shift in both academic research and industrial practice. While top-tier SE venues continue to show sustained or emerging focus on areas like automated testing and program repair, with researchers worldwide reporting continuous performance gains, the alignment of these academic advances with real industrial needs remains unclear. To bridge this gap, we first conduct a systematic analysis of 1,367 papers published in FSE, ASE, and ICSE between 2022 and 2025, identifying key research topics, commonly used benchmarks, industrial relevance, and open-source availability. We then carry out an empirical survey across 17 organizations, collecting 282 responses on six prominent topics, i.e., program analysis, automated testing, code generation/completion, issue resolution, pre-trained code models, and dependency management, through structured questionnaires. By contrasting academic capabilities with industrial feedback, we derive seven critical implications, highlighting under-addressed challenges in \emph{software requirements and architecture, the reliability and explainability of intelligent SE approaches, input assumptions in academic research, practical evaluation tensions, and ethical considerations}. This study aims to refocus academic attention on these important yet under-explored problems and to guide future SE research toward greater industrial impact.

\end{abstract}




\maketitle

\section{Introduction}
\label{sec:intro}

The emergence of large language models (LLMs), particularly since the widespread adoption of ChatGPT in late 2022, is significantly reshaping traditional academic research areas such as requirements analysis \cite{DBLP:journals/pacmse/LianWZLWZ25, DBLP:conf/re/LianMLZ24}, code completion and synthesis \cite{wang2024rlcoder,liu2024graphcoder}, automated testing \cite{jang2025unitcon, kim2025llamaresttest, bose2023columbus}, and software maintenance \cite{steenhoek2024dataflow, cao2024snopy, rahman2024towards}. Concurrently, AI-powered software engineering tools, exemplified by intelligent IDEs like GitHub Copilot \cite{GitHub-Copilot}, are being widely adopted by both novice and expert programmers across various domains.
Although academic research and industrial development are distinct domains, they remain deeply interconnected ~\cite{wang2023practitioners, lo2015practitioners, steenhoek2024closing}. Ideally, research questions and ultimate goals should originate from real-world industrial challenges, while academic outcomes should, in turn, drive industrial progress. To achieve this synergy, it is essential to periodically align academic research priorities and capabilities with the evolving needs of industry.

Several notable empirical studies have begun addressing the gap between academic research and practical industrial needs. For example, ResearchBot \cite{11050792} was designed to answer real-world questions from programming communities by retrieving and summarizing relevant academic literature. Szymon and Lech \cite{stradowski2023bridging} explored the application of machine learning-based software defect prediction in an industrial setting, specifically system-level testing for Nokia's 5G products. \citet{wang2023practitioners} investigated practitioners' expectations regarding automated code completion tools. \citet{steenhoek2024closing} conducted a user study to evaluate the AI-powered vulnerability detection and repair tool DeepVulGuard in a real development environment. A large-scale survey by Lo et al. ~\cite{lo2015practitioners} of 512 Microsoft engineers revealed that, despite 71\% of software engineering (SE) research (from 571 FSE/ICSE papers) being deemed essential or worthwhile, a significant misalignment with practitioners' needs persists. These studies, however, often focus on isolated tools or tasks. \emph{In the era of LLMs, a systematic reassessment of the alignment between the broader spectrum of SE research and evolving industrial needs is crucial. This paper examines the landscape of SE research in the AI era and evaluates its relevance to practical industry challenges.}

Our methodology begins with a systematic analysis of 1,367 recent publications from three premier SE conferences (ASE, FSE, and ICSE) from 2023 onward, following the global breakthrough of ChatGPT at the end of 2022. This analysis maps the statistical distribution of academic research hotspots and assesses the capabilities of state-of-the-art papers using public benchmarks as an anchor. Subsequently, we surveyed researchers in six prominent topics (each represented by over 40 publications) to understand current tool usage and identify unmet industrial needs. Based on 282 responses detailing technical limitations and expectations, we compared the academic state-of-the-art with practical industrial feedback. This comparison yielded seven key insights to inform future research directions.

Particularly, we address the following three research questions (RQs) in this study.

\begin{itemize}[leftmargin=1em]
    \item \textbf{RQ1: What are the dominant research themes and state-of-the-art capabilities in the three premier SE conferences}

     To answer RQ1, we collected all 1,367 publications from the research tracks of the three target conferences. Each publication was annotated based on several dimensions: its primary software lifecycle phase, research target, benchmarks and metrics used, evidence of industrial adoption, open-source availability, and repository popularity. Based on this annotation, we present the distribution of publications across software lifecycle phases and research topics. We then conduct a deep-dive analysis into 518 papers identified under six prominent research hotspots.
     
     We report six key insights here.\ding{172} \textbf{Testing Dominance}: Research on Testing remains the most prevalent topic over the three years, whereas Requirements was the least studied area.
    \ding{173} \textbf{Pervasiveness of AI}: Approximately 79.5\% of papers are relevant to intelligent techniques, spanning their integration with traditional methods, the development of new models, and associated ethical considerations.
    \ding{174} \textbf{Industrial Engagement}: A growing trend of industrial authorship is observed, indicating the increasing competitiveness of industry-led research in the AI era.
    \ding{175} \textbf{Benchmark Fragmentation}: Despite the popularity of benchmark construction, research is often evaluated against diverse benchmarks. This poses a significant challenge for the direct comparison of approaches, even within the same research topic.
    \ding{176} \textbf{Immature Capabilities}: Despite the rapid evolution of techniques, their practical capabilities remain largely immature. This is evident in the low performance observed in current research: for instance, the issue resolution rate on the SWE-Bench full suite is merely 19.31\%~\cite{ruan2024specrover}, while the defect detection precision of state-of-the-art methods stands at only about 20\% ~\cite{yin2025you}.
    \ding{177} \textbf{Focus on Functional Metrics}: Research in the six hotspots predominantly focuses on functional metrics (e.g., pass@k for code synthesis, BLEU for code summarization). More attention is needed on aspects such as required human effort in the post-processing stage, intellectual property, and private data protection.

    \item \textbf{RQ2: What are the prevailing automated technology needs in industry?}

    To address RQ2, we conducted a survey targeting industrial experts, focusing on the six most prominent academic research areas identified in RQ1: automated testing, program analysis, code generation and completion, automated issue resolution, software dependency management, and the underlying pre-trained code models. We distributed the survey through our professional networks and collected 282 responses from 17 industrial organizations after filtering out potentially invalid submissions (e.g., responses completed in under one minute). The respondent pool includes leading internet and software companies such as Intel, Microsoft, Huawei, and ByteDance; autonomous driving companies (such as Momenta); and five state key system engineering institutions in China. Each questionnaire was tailored to a specific research topic, soliciting insights on current usage of automated techniques, unmet industrial needs, and evaluations of relevant academic advancements. 
    
    Our analysis of the feedback reveals several key insights into the industrial landscape. \ding{172} \textbf{Automated Testing:} The adoption of automated testing tools is high (62.16\%), indicating the strong need of automated and advanced techniques. Practitioners express a need for more advanced intelligent techniques to enhance both effectiveness and efficiency, and they remain skeptical about the connection between technical novelty and practical applicability in complex real-world settings.
    \ding{173} \textbf{Program Analysis:} While the adoption of automated tools is high (90.0\%), with a clear trend toward AI/LLM-augmented traditional techniques, their usage hinges on user-centric features like automated fixes and workflow integration. Despite a steady stream of academic research, practitioners remain cautious due to the complexities of real-world applications.
    \ding{174} \textbf{Code Generation and Completion:} The most preferred tools are those deeply integrated into development workflows and native IDEs, such as GitHub Copilot and Cursor. Minimizing manual effort for code modification is a top priority, driving an urgent need for closer integration with project context and version-compatible APIs. The acceptance of academic techniques like RAG and multi-agent systems is conditional, hindered by the gap between research and practical usage.
    \ding{175} \textbf{Automated Issue Resolution:} Critical challenges impeding the routine use of automated techniques include a deficit of trust, high costs, poor integration into development workflows, and data security concerns. Industry expects these techniques to be evaluated in complex, real-world environments.
    \ding{176} \textbf{Software Dependency Management:} The top three industrial challenges are dependency conflicts, security vulnerabilities, and breaking changes. The most critically expected features are the ability to identify security vulnerability propagation paths and to facilitate compatible dependency updates.
    \ding{177} \textbf{Pre-trained Code Models:} Practitioner concerns are dominated by non-technical factors such as data security and legal assurance. There is also a clear industrial demand for addressing fundamental issues like computational performance and cost.

    \item \textbf{RQ3: Where do the most significant gaps exist between industrial needs and academic research?}

    By comparing academic capabilities with industrial feedback, we derive seven key implications to guide future research directions in the SE community.
    \ding{172} \textbf{Intelligent Software Requirements and Architecture:} Significant attention is needed in this under-explored area to bridge a critical gap in the software development lifecycle, yet it remains overlooked despite the paramount attention given to downstream tasks, such as code synthesis, issue resolution, and automated testing, which rely on it.
    \ding{173} \textbf{Reliability and Explainability:} These qualities are paramount for the successful industrial adoption of LLM-augmented SE approaches, yet remain a challenge in current research.
    \ding{174} \textbf{Divergent Test Inputs:} A key disconnect exists between academic test case generation (often from software requirements) and industrial practice (driven by testing requirements), limiting the applicability of research.
    \ding{175} \textbf{Cross-Language Program Analysis:} The proliferation of multi-language software systems creates a pressing need for analysis tools that transcend single-language boundaries.
    \ding{176} \textbf{Synthesis of Foundational Software:} The automated synthesis of high-assurance systems (e.g., OS kernels), which demand extreme functional correctness, robustness, and reliability, remains a largely open challenge.
    \ding{177} \textbf{Practically-Grounded Evaluation of Code Synthesis:} Research must evolve beyond standard benchmarks to incorporate evaluation metrics that reflect real-world usability and integration effort.
    \ding{178} \textbf{Ethical Considerations for Pre-trained Models:} Ethical aspects, including data security and privacy, are emerging as non-negotiable requirements for the deployment of pre-trained code models in industry.
    
\end{itemize}

\vspace{0.2em}
\textbf{Contributions.} The main contributions of this study are threefold.

\begin{itemize}[leftmargin=1.5em]
    \item To the best of our knowledge, this work presents the first comprehensive and in-depth comparison between academic capabilities and industrial needs in software engineering. From this comparison, we derive seven critical research directions to guide future academic efforts.
    \item We have conducted a systematic analysis of 1,367 publications from the top three SE conferences over the past three years. This analysis maps the landscape of academic research and benchmarks the capabilities within six prominent topics, yielding 11 key insights.
    \item We designed and distributed a topic-specific survey to investigate the adoption of automated tools and the perception of academic advancements in industry. Based on 282 qualified responses, we summarize concrete findings and industrial expectations for each research topic.
\end{itemize}

\vspace{0.2em}
\textbf{Paper Structure.}  The remainder of this paper is organized as follows. Section \ref{sec:academicAnalysis} presents our systematic analysis of academic research. Section \ref{sec:survey} details the methodology and findings of our industrial survey. In Section \ref{sec:implications}, we derive and discuss seven key implications by synthesizing academic capabilities with industrial feedback. Section \ref{sec:discussion} addresses threats to validity, followed by a discussion of the related work in Section \ref{sec:relatedWork}. Finally, Section \ref{sec:conclusion} provides the concluding remarks.

\section{Capability Analysis of Academic Research}
\label{sec:academicAnalysis}

To examine the Software Engineering community's research interests following the global breakthrough of ChatGPT, we gathered 1,367 papers from top-tier conferences, i.e., ASE, FSE, and ICSE. Our dataset includes all papers from the 2023–2025 editions of FSE and ICSE, as well as the 2023 and 2024 editions of ASE (the 2025 ASE papers were not yet published at the time of our study). We focused exclusively on the main research tracks, as they represent the most advanced academic contributions and typically contain fully developed studies (i.e., techniques and extensive evaluation).

For each paper, we manually annotated the following aspects based on the content of the paper and its assigned conference session:
\ding{172} \textbf{Classification by software lifecycle phase:} Each paper was categorized into one of six major software development phases: requirements, design, development, testing, maintenance, and management. This classification helps reveal the distribution of research focus across key development stages.
\ding{173} \textbf{Research topic:} 
    To clarify the academic tasks and objectives in the literature, we categorized the 1,367 papers based on their research goals, assigning each paper to one primary topic (e.g., Software Security \& Privacy) and one corresponding sub-topic (e.g., Malicious Code Detection) according to its ultimate goal, even though a paper could potentially relate to multiple topics.
    
   We conducted the topic annotation through an AI-augmented, paired human review process. First, DeepSeek-v3 generated initial labels from paper abstracts. Subsequently, four authors verified and corrected these annotations in a paired fashion.
    This manual refinement phase involved consolidating or removing certain topics to ensure coherence. For instance, we add the topic, \textit{Requirements Engineering and Software Architecture} beyond the auto-generated ones, to encompass sub-topics such as \textit{Requirements Elicitation and Analysis} and \textit{Architecture Design and Analysis}. Moreover, \textit{Compilation Testing} and \textit{Fuzz Testing} were consolidated into \textit{Automated Testing Technology} to enhance conceptual coherence.
    \noindent \textbf{\ding{174} Benchmarks and quantitative metrics:} We documented the benchmarks, evaluation metrics, and reported results to understand prevalent evaluation practices in top-tier research. This allows us to assess the applicability of these evaluations in real industrial settings.
    \textbf{\ding{175} Industrial adoption:} We noted whether the proposed methods have been adopted in industry, offering perspective on the practical impact and translational value of academic research.
  \textbf{\ding{176} Presence of industrial authors:} We identified papers with industry-affiliated authors and compared their research themes with those from purely academic affiliations. This helps highlight industry-relevant topics and potential gaps between academic and industrial research priorities.
  \textbf{\ding{177} Open-source availability and repository popularity:} We recorded whether the research artifacts are open-source and the number of stars received by their repositories. This serves as an indicator of reproducibility, practical utility, and community recognition.

\begin{figure}[ht]
    \centering
    \begin{tikzpicture}
        \begin{axis}[
            width=10cm,
            height=5cm,
            ybar,
            bar width=6pt,
            symbolic x coords={2023,2024,2025},
            xtick=data,
            ymin=0, ymax=50,
            ytick={0,10,20,30,40,50},
            xlabel={Years},
            ylabel={Percentage (\%)},
            grid=major,
            enlarge x limits={0.15},
            legend style={
                at={(0.5,1.02)},
                anchor=south,
                legend columns=6,
                font=\footnotesize,
                /tikz/every even column/.append style={column sep=0.1cm},
                draw=none,
                fill=none
            },
            legend image code/.code={
                \draw[#1,fill] (0cm,-0.1cm) rectangle (0.3cm,0.1cm);
            },
            nodes near coords,
            nodes near coords align={vertical},
            nodes near coords style={font=\tiny},
        ]

        \addplot[brown, fill=brown!40] coordinates {(2023,1.2) (2024,0.7) (2025,1.5)};
        \addlegendentry{Req}
        \addplot[orange, fill=orange!50] coordinates {(2023,1.2) (2024,2.1) (2025,1.5)};
        \addlegendentry{Design}
        \addplot[teal!60!black!40, fill=teal!60!black!40] coordinates {(2023,20.7) (2024,21.3) (2025,30.4)};
        \addlegendentry{Dev}
        \addplot[blue, fill=blue!50] coordinates {(2023,33.0) (2024,36.5) (2025,45.5)};
        \addlegendentry{Test}
        \addplot[violet, fill=violet!30] coordinates {(2023,22.1) (2024,19.8) (2025,17.3)};
        \addlegendentry{Maint}
        \addplot[red, fill=red!50] coordinates {(2023,4.4) (2024,3.8) (2025,2.0)};
        \addlegendentry{Mgmt}

        \end{axis}
    \end{tikzpicture}
    \caption{Percentage of Papers across Software Lifecycle Phases (2023–2025)}
    \label{fig:bar_chart}
\end{figure}
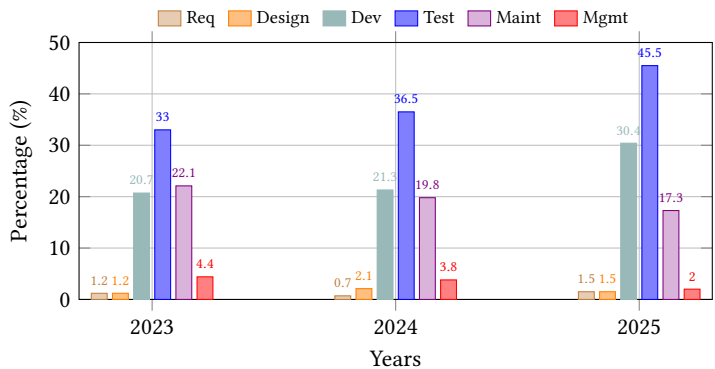

Figure \ref{fig:bar_chart} illustrates the distribution across phases, whereas Table \ref{tab:topic-classification} presents the frequency and proportion of papers across topics and sub-topics. The predominant categories are indicated in bold. Consistent with our coding scheme, each paper contributed to only one count in these distributions.

\definecolor{cat1}{RGB}{198, 219, 239}
\definecolor{cat2}{RGB}{255, 205, 210}
\definecolor{cat3}{RGB}{197, 225, 165}
\definecolor{cat4}{RGB}{255, 224, 178}
\definecolor{cat5}{RGB}{179, 157, 219}
\definecolor{cat6}{RGB}{255, 249, 196}
\definecolor{cat7}{RGB}{178, 235, 242}
\definecolor{cat8}{RGB}{255, 204, 204}
\definecolor{cat9}{RGB}{255, 224, 229}
\definecolor{cat10}{RGB}{232, 234, 246}
\definecolor{cat11}{RGB}{220, 237, 200}

\begin{footnotesize}
\begin{longtable}{|p{5.5cm}|p{7cm}|p{2cm}|}
\caption{Papers Distribution by Topics} \label{tab:topic-classification} \\
\hline
\textbf{Topics} & \textbf{Sub-topics} & \textbf{Count \& Ratio} \\
\hline
\endfirsthead
\multicolumn{3}{c}%
{\tablename\ \thetable\ -- \textit{Continued from previous page}} \\
\hline
\textbf{Topics} & \textbf{Sub-topics} & \textbf{Count \& Ratio} \\
\hline
\endhead
\hline \multicolumn{3}{r}{\textit{Continued on next page}} \\
\endfoot
\hline
\endlastfoot

\rowcolor{cat1} Anomaly Detection \& System Fault Diagnosis &\textbf{Cloud Service Root Cause Analysis} & \textbf{19 (1.4\%)}\\
\rowcolor{cat1} (60, 4.4\%) & Exception Detection \& Fault Localization in Microservices Systems & 14 (1.0\%)\\
\rowcolor{cat1} & Log Anomaly Detection & 13 (1.0\%)\\
\rowcolor{cat1} & Log Parsing & 14 (1.0\%)\\

\rowcolor{cat2} Code Generation \& Understanding & \textbf{Code Generation \& Completion} & \textbf{61 (4.5\%)}\\
\rowcolor{cat2} (147, 10.8\%) & Code Refactoring \& Smell Detection & 20 (1.5\%)\\
\rowcolor{cat2} & Code Search \& Recommendation & 14 (1.0\%)\\
\rowcolor{cat2} & Code Summarization \& Comment Generation & 23 (1.7\%)\\
\rowcolor{cat2} & Code Translation \& Migration & 20 (1.5\%)\\
\rowcolor{cat2} & Type Inference \& Code Normalization & 9 (0.7\%)\\

\rowcolor{cat3} Cross-Domain \& Emerging Topics & AI Ethics \& Fairness & 24 (1.8\%)\\
\rowcolor{cat3} (63, 4.6\%) & \textbf{Autonomous Driving System Testing} & \textbf{27 (2.0\%)}\\
\rowcolor{cat3} & Neuroscience \& Programming Cognition & 5 (0.4\%)\\
\rowcolor{cat3} & Quantum Software Engineering & 5 (0.4\%)\\
\rowcolor{cat3} & Sustainable Software Engineering & 2 (0.1\%)\\

\rowcolor{cat4} Formal Methods \& Verification & Compiler Correctness Verification & 5 (0.4\%)\\
\rowcolor{cat4} (54, 4.0\%) & Loop Invariant Inference & 4 (0.3\%)\\
\rowcolor{cat4} & Model Checking \& Symbolic Execution & 12 (0.9\%)\\
\rowcolor{cat4} & \textbf{Program Verification \& Theorem Proving} & \textbf{25 (1.8\%)}\\
\rowcolor{cat4} & Requirements Formal Modeling \& Verification & 4 (0.3\%)\\
\rowcolor{cat4} & Temporal Logic \& LTL Games & 4 (0.3\%)\\

\rowcolor{cat5} HCI \& Developer Tools &  Debugger \& Test Tool Optimization & 8 (0.6\%)\\
\rowcolor{cat5} (50, 3.7\%) & \textbf{Developer Experience \& Flow Analysis} & \textbf{24 (1.8\%)}\\
\rowcolor{cat5} & Development of IDE Tools & 9 (0.7\%)\\
\rowcolor{cat5} & Visualization \& Cognitive Support & 9 (0.7\%)\\

\rowcolor{cat6} \textbf{Program Analysis \& Software Testing} & \textbf{Automated Testing Technology} & \textbf{211 (15.4\%)}\\
\rowcolor{cat6} \textbf{(399, 29.2\%)} & Program Analysis & 120 (8.8\%)\\
\rowcolor{cat6} & Quantum/Deep Learning Testing & 39 (2.9\%)\\
\rowcolor{cat6} & Test Efficiency Optimization & 21 (1.5\%)\\
\rowcolor{cat6} & Test Result Processing & 8 (0.6\%)\\

\rowcolor{cat7} Requirements Engineering \& Software Architecture & Architecture Design \& Analysis & 8 (0.6\%)\\
\rowcolor{cat7} (32, 2.3\%) & Architecture Evolution \& Maintenance & 2 (0.1\%)\\
\rowcolor{cat7} & Requirements Elicitation \& Analysis & 9 (0.7\%)\\
\rowcolor{cat7} & Requirements Traceability \& Linkage & 3 (0.2\%)\\
\rowcolor{cat7} & \textbf{Software Architecture Recovery \& Refactoring} & \textbf{10 (0.7\%)}\\

\rowcolor{cat8} SE for AI & AI-Generated Content Identification \& Hallucination Detection & 7 (0.5\%)\\
\rowcolor{cat8} (97, 7.1\%) & AI System Integration \& Reproduction & 2 (0.1\%)\\
\rowcolor{cat8} & ML-based Text Mining \& Information Extraction & 9 (0.7\%)\\
\rowcolor{cat8} & Model Robustness  & 31 (2.3\%)\\
\rowcolor{cat8} & \textbf{Pre-trained Code Models} & \textbf{48 (3.5\%)}\\

\rowcolor{cat9} Software Maintenance \& Evolution & Automated Code Review & 15 (1.1\%)\\
\rowcolor{cat9} (260, 19.0\%) & \textbf{Automated Program Repair \& Issue Resolution} & \textbf{126 (9.2\%)}\\
\rowcolor{cat9} & Code Clone Detection & 14 (1.0\%)\\
\rowcolor{cat9} & Developer Behavior Analysis & 28 (2.0\%)\\
\rowcolor{cat9} & Open Source Community \& Collaboration Models & 19 (1.4\%)\\
\rowcolor{cat9} & Software Dependency Management & 41 (3.0\%)\\
\rowcolor{cat9} & Sustainability of Open Source Projects & 17 (1.2\%)\\

\rowcolor{cat10} Software Security \& Privacy & Binary Code Analysis & 29 (2.1\%)\\
\rowcolor{cat10} (157, 11.5\%) & Code Poison/Backdoor Detection & 7 (0.5\%)\\
\rowcolor{cat10} & Exploit Path Generation & 3 (0.2\%)\\
\rowcolor{cat10} & Information Leakage Assessment & 9 (0.7\%)\\
\rowcolor{cat10} & Malicious Code Detection & 21 (1.5\%)\\
\rowcolor{cat10} & \textbf{Smart Contract Security} & \textbf{47 (3.4\%)}\\
\rowcolor{cat10} & User Security \& Ethics Compliance & 29 (2.1\%)\\
\rowcolor{cat10} & Vulnerability Prediction \& Synthesis & 12 (0.9\%)\\

\rowcolor{cat11} Systems \& Performance Engineering & \textbf{Performance Analysis \& Optimization} & \textbf{34 (2.5\%)}\\
\rowcolor{cat11} (48, 3.5\%) & Resource Leak Detection & 7 (0.5\%)\\
\rowcolor{cat11} & Software Construction \& Deployment & 7 (0.5\%)\\
\end{longtable}
\end{footnotesize}

Based on the information we collect, the statistical results are as follows:
\begin{itemize}[leftmargin=1em]
    \item As shown in Fig. ~\ref{fig:bar_chart}, \textbf{papers on testing remained the most prevalent topic over the three-year period (588 papers, 43\% on average), showing a consistent upward trend. In contrast, requirements was the least studied area (17 papers, 1.2\% on average)}. A similar growth trend was observed for development topics, such as code synthesis. This disparity indicates that contemporary SE research prioritizes implementation and quality assurance over earlier lifecycle phases like requirements analysis and software design.

    \item \textbf{About 79.5\% papers are relevant to artificial intelligence (AI) technologies},  primarily those involving deep learning techniques like large language models, with a smaller number utilizing traditional machine learning techniques. 
    This indicates that the majority of recent software engineering research actively incorporates or investigates intelligent technologies. In addition, we observe that research related to intelligent technologies appears across the all six software lifecycle phases.

    \item \textbf{Besides traditional SE tasks like testing and code maintenance, new topics emerge with the wider and deeper application of DL.} Traditional such as Program Analysis \& Software Testing (29.1\%), Software Maintenance \& Evolution (19.6\%), and Software Security \& Privacy (11.4\%) remain central topics within the field, while the research of SE4AI (like their modularity \cite{10172675}) are emerging and witnessing growing impact. Smaller yet strategically significant areas, including cross-domain and socio-technical studies ~\cite{10.1109/ICSE55347.2025.00070, 10298485, 10.1109/ICSE48619.2023.00133}, also indicate the growing diversification and the expansion of software engineering into safety-critical, ethical, and sustainable contexts.

    \item \textbf{Of the papers studied, an average of 26.8\% over the three years reported research that was applied in industry, and 16.2\% included authors affiliated with industrial organizations.} Furthermore, the proportion of industrial authors was higher in 2025 (19.9\%) than in 2024 (12\%). This suggests a growing emphasis among researchers on conducting practically relevant studies. One contributing factor is the field's reliance on computational resources, which are largely controlled by industry. 

    \item \textbf{Although most studies release a replication package and 32.9\% have a GitHub repository, few repositories achieve significant influence.} For instance, of the 449 GitHub repositories, only three have gained 1,000+ stars ~\cite{11029883, ruan2024specrover, 10.1145/3729355}, and merely one exceeds 5,000 stars ~\cite{11029883}.

\end{itemize}

Our assessment of academic research capabilities focused on the six most prominent topics in Table \ref{tab:topic-classification}, each with over 40 papers. Smart Contract Security was excluded on the basis that it represents a specific application, not a classical SE task, despite having 47 relevant papers. Our microscopic analysis focused on the benchmarks and evaluations used in these studies. Given the diversity of approaches, we employed common, public benchmarks as a consistent basis for comparison; consequently, only research results utilizing these benchmarks are included. All evaluation data was sourced directly from the original publications without re-evaluation.

\subsection{Automated Testing Technology (211 papers, 15.4\%):} 
\label{subsec:automatedTesting}

    \subsubsection{Common benchmarks and the corresponding approaches' capabilities.}
    In the domain of automated software testing, some benchmarks have been established and widely used to evaluate various approaches and tools. These benchmarks primarily encompass three categories: \textbf{unit testing, mobile application testing, and REST API}. Each benchmark provides real-world or synthetic datasets with known bugs, enabling the systematic comparison of key metrics such as code coverage, fault detection, and scalability across different techniques. We summarize the common benchmarks, the latest approaches, and evaluation results in Table ~\ref{tab:autotest}.

     \textbf{At the unit-testing level, datasets such as Defects4J ~\cite{just2014defects4j} and Bears ~\cite{madeiral2019bears} are widely employed. These benchmarks contain hundreds of real-world projects and historical bug data, serving as a foundation for evaluating the effectiveness of automated bug reproduction and unit testing.} The representative publications on these benchmarks are listed here.
     \begin{itemize}[leftmargin=1.7em]
         \item The approach of \texttt{UnitCon} ~\cite{jang2025unitcon} successfully generated runtime exception-triggering test cases for 52.5\% of the open-source Java projects in Defects4J and Bears within ten minutes, achieving 1.2–3.6 times more than the baselines. 
         
         \item Towards null pointer exceptions (NPEs) identification, 
        \texttt{UnitCon} discovered 21 previously unknown NPE bugs, 15 of which were later confirmed and fixed by developers, demonstrating its practical impact on real-world projects.  And the Null Pointer Exception-guided testing strategy (\texttt{NpeTest}) ~\cite{lee2024effective} achieved a 78.9\% reproduction rate on 108 known NPEs collected from 96 real-world projects, which were drawn from  Defects4J and Bears, representing a 38.7\% improvement over the baseline. 
         Moreover, \texttt{NpeTest} uncovered 89 previously unknown NPEs in industrial projects, which is 7 to 37 more than the baselines.

         \item The approach of \texttt{AUGER} ~\cite{yin2025you} successfully triggered 84 defects for the 723 method-level defects in Defects4J, achieving a recall of 12\%. Compared with the baselines, \texttt{AUGER} triggered more defects (84 vs 23). 
         For effectiveness in real-world projects, \texttt{AUGER} successfully triggered 6 defects out of 41 method-level defects collected from high-quality open-source projects, achieving a recall of 15\%. Although this result was better than the baselines, \textbf{\emph{we believe it still requires substantial improvement for practical use}}, since 88\% of the method-level defects in Defects4J and 85\% of the real-world method-level defects were not triggered, and \texttt{AUGER} performs poorly on multi-hunk defects involving multiple methods.
     \end{itemize}

    \textbf{For mobile application testing, AndroTest ~\cite{choudhary2015automated}, Themis ~\cite{su2021benchmarking} and F-Droid ~\cite{fdroid2024} are the most commonly used benchmarks.} 
    \begin{itemize}[leftmargin=1.7em]
        \item \textbf{AndroTest provides a set of open-source Android applications designed for evaluating coverage-guided and crash-oriented test generation methods.} For instance, the approach of \texttt{COLUMBUS} ~\cite{bose2023columbus} achieved 58\% code coverage, which is 0.05-0.31 higher than baselines, and discovered 137 crashes in the AndroTest, representing 1.23-5.48 times more crashes than baselines. Besides, \texttt{COLUMBUS} is able to find 70 crashes in 54 out of 140 real-world Play Store apps, demonstrating its strong fault exploration ability in real-world apps.  Similarly, \texttt{DQT} ~\cite{lan2024deeply}, evaluated on 22 apps from AndroTest and 8 open-source apps from GitHub, achieved an average instruction coverage (i.e., the coverage of software requirements or user stories by test cases) of 46.8\%, which is 0.05 to 0.18 higher than baselines, and detected an average of 5.47 unique faults per app, exceeding baselines by 1.6 to 3.84 faults.

        \item \textbf{The Themis benchmark includes Android applications with genuine defects, providing a more realistic evaluation environment. And the F-Droid repository serves as another valuable source of real-world Android applications for empirical studies.} The representative approaches include \texttt{BugHunter} ~\cite{chen2025standing}, which achieved a defect recall of 0.64 with 0.53 activity coverage (the percentage of executed activities or operational paths) and 0.52 code coverage on Themis and F-Droid. Compared to baselines, this represents an improvement of 0.24–0.40 in recall, 0.02–0.18 in activity coverage, and 0.05–0.19 in code coverage. Furthermore, \texttt{BugHunter} has identified 49 previously unknown crash bugs in real-world applications from the Google Play store. Of these, 33 have been promptly addressed by the developers, 9 have been confirmed. Only 7 of these new bugs are detected by the best baseline. Another  approach \texttt{GPTDroid} ~\cite{liu2024make}, leveraging large language models, achieved 66\% recall with 75\% activity coverage and 66\% code coverage across 93 apps from Themis and Google Play. It outperforms the best baseline by achieving over 32\% higher activity coverage, 20\% higher code coverage, and detecting 31\% more bugs. Moreover, \texttt{GPTDroid} identify 53 new bugs on Google Play, of which 35 have been confirmed or fixed.
    \end{itemize}

    \textbf{In service-level testing, REST API benchmarks have become an important evaluation setting. Common benchmark suites include various RESTful services like Features Service, LanguageTool, and REST Countries, offering both open-source and real-world APIs for systematic testing.} The recent approaches mostly are based on reinforcement learning and large language models.

    Reinforcement learning-based testing, such as \texttt{ARAT-RL} ~\cite{kim2023adaptive}, attained 36.3\% branch coverage, 58.5\% line coverage, and 59.4\% method coverage across ten RESTful services.
    Similarly, the LLM-driven approach \texttt{LlamaRestTest} ~\cite{kim2025llamaresttest}, evaluated on 12 real-world RESTful services, achieved 28.3\% branch coverage, 55.3\% line coverage and 55.8\% method coverage, \emph{exceeding \texttt{ARAT-RL} by 9.2, 11.9 and 14.1 percentage points respectively}. It also detected 204 internal server errors across the evaluated REST APIs, which is 44 more than \texttt{ARAT-RL}. \texttt{AutoRestTest} ~\cite{kim2024multi}, evaluated on the same 12 real-world RESTful services as LlamaRestTest, achieved 32.1\% branch coverage, 58.3\% line coverage and 58.3\% method coverage, exceeding \texttt{ARAT-RL} by 12.3, 14.3 and 16.2 percentage points respectively. It also detected 42 internal server errors across the evaluated REST APIs, which is 11 more than \texttt{ARAT-RL}. The latter two studies demonstrate significant improvements over ARAT-RL, with AutoRestTest achieving more coverage gains and LlamaRestTest detecting more internal server errors.

    \begin{table}[h]
    \centering
    \small
    \caption{Performance of Representative Methods on Common Testing Benchmarks}
    \label{tab:autotest}
    \begin{tabular}{lcccc}
    \toprule
    \textbf{Dataset} & \textbf{Method} & \textbf{Coverage \& Recall} & \textbf{Detection of Real-World Defects} \\
    \midrule
    Defects4J+Bears~\cite{madeiral2019bears}, etc. & UnitCon~\cite{jang2025unitcon} & Recall: 0.53 & 21 new NPE bugs \\
    Defects4J+Bears, etc. & NPE-Test~\cite{lee2024effective} & Recall: 0.79 & 89 new NPEs \\
    Defects4J~\cite{just2014defects4j} & AUGER~\cite{yin2025you} & Recall: 0.12 &  6 method-level defects\\
    \hline
    AndroTest~\cite{choudhary2015automated} & COLUMBUS~\cite{bose2023columbus} & Code Coverage: 58\% & 70 crashes \\
    AndroTest+other apps & DQT~\cite{lan2024deeply} & Instruction Coverage: 46.8\% & 5.47 faults/app \\
    Themis~\cite{su2021benchmarking}+F-Droid~\cite{fdroid2024} & BugHunter~\cite{chen2025standing} & Activity Coverage: 53\%, Recall: 0.64 & 49 new crash bugs \\
    Themis+other apps & GPTDroid~\cite{liu2024make} & Act. Cov.: 75\%, Code Cov.: 66\%, Recall: 0.66 & 53 new bugs on Google Play \\
    \bottomrule
    \end{tabular}
\end{table}

    \subsubsection{Expanding Scope: Approaches Beyond Standard Benchmarks.}
    \textbf{\emph{In addition to evaluations on public benchmarks, a substantial segment of testing research utilizes custom datasets, often constructed from sources like GitHub or the Google Play Store.}} The tasks are usually diverse, and we introduce the relevant studies here. 

    \begin{itemize}[leftmargin=1.7em]
        \item \texttt{Libro} ~\cite{kang2023large} \emph{generates test cases} to reproduce real software failures. It successfully reproduced 32.2\% (10/31) of the bugs in the GitHub Recent Bugs dataset~\cite{kang2023large}, a notable result that underscores the inherent difficulty of this task. 
        \item \texttt{MUT} ~\cite{gao2024mut} focuses on \emph{migrating unit tests} from Java to C++. On a benchmark of 15 open-source projects, it successfully migrated and integrated 253 out of 550 tests (46\%) into the target projects, uncovering 7 previously unknown defects in the process. 
        \item \texttt{LLMDroid} ~\cite{wang2025llmdroid} enhances \emph{GUI test automation}, demonstrating its effectiveness on 14 popular apps from Google Play. It achieved average improvements of 26.16\% in code coverage and 29.31\% in activity coverage over the baseline tools like DroidBot ~\cite{li2017droidbot} and FastBot ~\cite{lv2022fastbot2}. Similarly, QTypist ~\cite{liu2023fill} automates \emph{the generation of meaningful text inputs} for mobile testing, achieving an 87\% input success rate on a benchmark of 106 Android apps from Google Play. 
    \end{itemize}

    The field is also expanding to cover \textbf{\emph{a wider range of systems and programming languages}}. For example, \texttt{Rust-twins} ~\cite{yang2024rust} targets \emph{compiler testing} for Rust. It achieved twice the code coverage of its predecessor, \texttt{RustSmith}~\cite{sharma2023rustsmith}, and during a 24-hour run, it triggered 10 compiler crashes and 229 inconsistencies, leading to the discovery of 12 new bugs (8 confirmed or fixed). \texttt{Mozi} ~\cite{liang2024mozi} focuses on finding correctness and performance bugs in \emph{database management systems} (MySQL, PostgreSQL, etc.). Over three weeks, it discovered 101 new bugs, with 90 confirmed and 57 fixed by developers.

\vspace{1mm}
\begin{custommdframed}
\textbf{Finding 1 (Academic Automated Testing Technology):} The commonly used metrics focus on the coverage of requirements (e.g., instruction coverage), code (e.g., activity, line and method) and defect detection rate. However, from the practical usage perspective, \textbf{the readability, maintainability, and reliability of test cases generated by automated testing tools still require further attention}~\cite{nie2023learning}.
\end{custommdframed}
\vspace{1mm}

\vspace{1mm}
\begin{custommdframed}
\textbf{Finding 2 (Academic Automated Testing Technology):} Although automated testing approaches are continuously improved, their effectiveness remains limited, as evaluated by requirement coverage (nearly 50\%~\cite{bose2023columbus,lan2024deeply,chen2025standing}) and defect detection precision (around 20\%~\cite{yin2025you}). \textbf{This indicates that significant human effort is still required to use these approaches in practice.} 
\end{custommdframed}
\vspace{1mm}

    \subsection{Program Analysis (120 papers, 8.8\%):} 
    \label{subsec: programAnalysis}
    
    \subsubsection{Common benchmarks and the corresponding approaches’ capabilities.} A variety of benchmarks have been developed to to support the evaluation of static or dynamic analysis techniques across different programming languages and application domains. \textbf{This research domain encompasses two primary foci: General Program Analysis and Mobile App Analysis. 
The former approach typically analyzes programs without constraints on system types, employing established benchmarks such as 
\emph{SV-COMP} and \emph{SPEC CPU}~\cite{bucek2018spec} for C/C++, \emph{DaCapo}~\cite{blackburn2006dacapo} for Java, 
and \emph{BugsInPy}~\cite{widyasari2020bugsinpy} for Python. The latter specialization utilizes mobile-specific benchmarks including 
\emph{DroidBench}~\cite{droidbench2023}, \emph{ICC-Bench}~\cite{icc-bench}, and \emph{F-Droid}~\cite{fdroid2024}.}

    We collect the performance evaluation of the related methods on the common benchmarks from their original papers, and show the results on \textit{General Program Analysis} in Table~\ref{tab:generalbench}, and the performance on \textit{Mobile App Analysis} is presented in Table~\ref{tab:mobile}.

    \begin{itemize}[leftmargin=1.7em]
    	\item \textbf{The first zone is about SV-COMP ~\cite{SVCOMP2023}, which is a benchmark suite originating from the annual \textit{Software Verification Competition} \footnote{\url{https://sv-comp.sosy-lab.org/2025/}}, designed to evaluate C program verification tools' capability on various verification tasks such as \textit{overflow detection}, \textit{reachability analysis}, and \textit{memory safety checking}}. SV-COMP has been used by multiple academic study, such as \emph{Parf}~\cite{wang2024parf}, \emph{PBEAR}~\cite{kim2024pbe}, and \emph{Laplace}~\cite{lee2023statistical}. Particularly, \texttt{Parf} achieved a 33.3\% verification accuracy in the \emph{NoOverflow task}, exceeding the baseline by 1.1 percentage points. While \texttt{PBEAR} reached a 93.8\% success rate in \emph{ReachSafety property falsification tasks} (a category of verification tasks aimed at determining whether an error state in a program is reachable) and \emph{the memory usage} of the baselines ranged from 1.24x to 3.27x that of \texttt{PBEAR}. The \texttt{Laplace estimator} successfully estimated \emph{the reaching probability} for all 32 programs in a short period of time generally, while the baselines failed to estimate the accurate reaching probability of nearly half of the programs. 
    	
    	\textbf{The SPEC CPU benchmark suite ~\cite{speccpu2006} is another commonly employed benchmark for assessing C/C++ program performance as well as analysis overhead.} For example, on this benchmark, \texttt{Light-Fusion} ~\cite{wang2025boosting} demonstrates remarkable efficiency by identifying approximately 80\% of redundant summaries. Compared to the baseline, it achieves a 45\% reduction in time overhead and a 27\% reduction in memory overhead, delivering an average performance gain of 632.1 times. Another approach \texttt{OLASan} ~\cite{wang2025practical} demonstrates its effectiveness in optimizing memory access checks by achieving a significant reduction in runtime overhead—6.52\% to 51.18\% lower than the baselines.

    	\textbf{The DaCapo benchmark suite ~\cite{blackburn2006dacapo}, which includes representative open-source Java applications such as Eclipse, Batik, and Tomcat, is widely used to evaluate the effectiveness and efficiency of program analysis.} For example, the proposed \texttt{s-k-obj} approach ~\cite{lu2023automatic} automatically \emph{summarizes libraries}, achieving a 2.1×–2.3× speedup over the baselines while maintaining similar or better precision on it. \texttt{Galette} ~\cite{hough2025dynamic}, an approach for \emph{dynamic taint tracking} in the Java virtual machine, achieves a mean execution time overhead of 7.393x (8.420 and 5593.692 for the baselines) and a mean memory overhead of 1.536x (1.379 and 9.971 for the baselines). \texttt{MPA} ~\cite{li2025module}, a module-aware \emph{context-sensitive pointer analysis} method, significantly improves pointer analysis precision (up to 97.5\%) while maintaining comparable runtime to conventional context-insensitivity analysis.

    	\item  \textbf{BugsInPy ~\cite{widyasari2020bugsinpy} offers \emph{over 1,800 real-world Python programs} for fault localization and behavioral prediction tasks.} Based on it, the \texttt{PREDEX} ~\cite{li2025blended} model \emph{predicts execution traces} with the metrics of ETA-F, ETA-P, and ETA-S scores (where ETA-F measures full execution trace prediction accuracy, ETA-P measures the proportion of correctly predicted prefix traces before the first error, and ETA-S measures next-statement prediction accuracy) of 50.7\%, 66.1\%, and 75.4\%, respectively, achieving absolute improvements of 16.7\%–21.7\%, 13.4\%–20.4\%, and 3.4\%–16.6\% over the baselines. The lightweight \texttt{gumtree-simple} ~\cite{falleri2024fine} tool improves edit script generation efficiency, reducing runtime from 205.65s to 4.07s and shortening 74\% of \emph{generated edit scripts} compared to \texttt{GumTree} ~\cite{falleri2014fine} (a widely used open-source heuristic that compares two abstract syntax trees (ASTs) to generate edit scripts describing their structural differences).
    	\textbf{\emph{The above content reveals that, despite employing the same benchmark suites, the relevant approaches pursue distinct aims and employ heterogeneous evaluation metrics.}}
    	
    	\item  \textbf{Benchmark studies on \emph{DroidBench}, \emph{ICC-Bench}, and \emph{F-Droid} underscore significant improvements in \emph{Android} static analysis.} In \emph{vulnerability detection}, \texttt{PacDroid} ~\cite{chen2025pacdroid} achieves an F-measure of 90\%, substantially outperforming baseline methods (65\%–77\%). For \emph{taint analysis}, \texttt{ViaLin} ~\cite{ahmed2023vialin} not only attains an F-measure of 94.9\% but also identifies 92.3\% of true-positive paths, far exceeding \texttt{FlowDroid}'s ~\cite{arzt2014flowdroid} recall of 65.4\%. Within the IFDS framework (an popular interprocedural data-flow framework used for solving context-sensitive static analysis problems), analysis efficiency is also enhanced: \texttt{MERGEDROID} ~\cite{gui2023merge} and \texttt{DStream} ~\cite{wang2023dstream} report speedups of 9× and 7.11× over \texttt{FlowDroid} on F-Droid while preserving precision.
    \end{itemize}

    \begin{table}[!htbp]
    \centering
    \caption{Performance on General Program Analysis Benchmarks (C, Java, and Python)}
    \label{tab:generalbench}
    \resizebox{1.0\textwidth}{!}{
    \begin{tabular}{lccc}
    \toprule
    \textbf{Benchmark} & \textbf{Method} & \textbf{Task} & \textbf{Main Results} \\
    \midrule
    SV-COMP 2022 (C)~\cite{SVCOMP2023} & Parf ~\cite{wang2024parf} & NoOverflow Verification & 33.3\% verification accuracy \\
    SV-COMP 2022 (C) & PBEAR ~\cite{kim2024pbe} & ReachSafety Falsification & 93.8\% success rate, less memory usage \\ 
    SV-COMP 2021 (C) & Laplace ~\cite{lee2023statistical} & Reachability Estimation & Reachability estimated: Full (100\%) \\ \hline
    SPEC CPU 2017 (C)~\cite{speccpu2006} & Light-Fusion~\cite{wang2025boosting} & Path-sensitive Value Flow Analysis & Redundancy identified: 79.9\%, 
speedup: 632.1x \\
    SPEC CPU 2006/17 (C) & OLASan~\cite{wang2025practical} & Memory Safety Detection & 6.52–51.18\% lower runtime than baselines\\ \hline
    DaCapo (Java)~\cite{blackburn2006dacapo} & s-k-obj~\cite{lu2023automatic} & Library Summary Generation & 2.1–2.3× speedup, same or better precision \\
    DaCapo (Java) & Galette~\cite{hough2025dynamic} & Dynamic Taint Tracking & 7.39× time overhead, 1.54× memory overhead \\
    DaCapo (Java) & MPA~\cite{li2025module} & Pointer Analysis & 97.5\% precision, near context-insensitive speed \\ \hline
    BugsInPy (Python)~\cite{widyasari2020bugsinpy} & PREDEX~\cite{li2025blended} & Execution Trace Prediction & ETA-F/P/S: 50.7\%, 66.1\%, 75.4\% \\
    BugsInPy (Python) & gumtree-simple~\cite{falleri2024fine} & Edit Script Generation & 74\% shorter scripts, 50× faster than GumTree~\cite{falleri2014fine} \\
    \bottomrule
    \end{tabular}
    }
    \end{table}

    \begin{table}[h]
    \centering
    \small
    \caption{Performance on Mobile App Analysis Benchmarks}
    \label{tab:mobile}
    \begin{tabular}{lccc}
    \toprule
    \textbf{Benchmark} & \textbf{Method} & \textbf{Task} & \textbf{Main Results} \\
    \midrule
    DroidBench~\cite{droidbench2023}+ICC-Bench~\cite{icc-bench} & PacDroid~\cite{chen2025pacdroid} & Vulnerability Detection & R:93\%, P:88\% and F1:90\%\\
    DroidBench+ICC-Bench & ViaLin~\cite{ahmed2023vialin} & Taint Flow Detection & R:92.3\%, P:97.6\% and F1: 94.9\% \\ \hline
    F-Droid~\cite{fdroid2024} & MERGEDROID~\cite{gui2023merge} & IFDS Taint Analysis & 9× faster than FlowDroid\\
    F-Droid & DStream~\cite{wang2023dstream} & Large-scale IFDS Analysis & 7.1× faster than FlowDroid\\
    \bottomrule
    \end{tabular}
    \end{table}

    \subsubsection{Expanding Scope: Approaches Beyond Standard Benchmarks.} 
    Besides the above common benchmarks, \textbf{\emph{researchers are constructing more complex benchmarks covering more programming languages and diverse tasks.}} For instance, \texttt{Lazifier} ~\cite{turcotte2023increasing} \emph{refactors JavaScript applications} and, in its evaluation on a benchmark of 10 open-source front-end applications, achieved an average reduction of 36.2\% in initial size, and improved initial loading time by 29.7\%. \texttt{OOM-Guard} ~\cite{chen2023oom}, designed to \emph{handle memory allocation failures}, was assessed on two Rust system projects, Bento-fs and rCore, achieving memory overhead below 10\% and execution time overheads of 3\% and 5\%, respectively, while reducing developers' code modifications for out of memory handling. \texttt{TIGER} ~\cite{wang2024tiger}, a \emph{Python type inference} tool, was tested on the ManyTypes4Py benchmark ~\cite{mir2021manytypes4py}—a collection of type-annotated Python functions from 5,996 GitHub projects—achieving an overall Top-5 exact match accuracy of 94.6\% in the task of Python type inference. \texttt{VD-Guard} ~\cite{liu2023vd}, targeting \emph{Direct Memory Access (DMA)-related vulnerabilities}, successfully detected all 10 known vulnerabilities in a dataset of a widely used open-source machine emulator and virtualizer (QEMU) ~\cite{bellard2005qemu} virtual device flaws.

    \textbf{Overall, these benchmarks together provide comprehensive evaluation settings for diverse program analysis techniques across multiple languages and domains, enabling precise comparisons in performance, scalability, and analysis accuracy.} In principle, the use of common benchmarks should allow for direct comparison of different approaches; however, as shown in Table~\ref{tab:generalbench} and~\ref{tab:mobile}, \textbf{\emph{even when the same benchmarks are used, the specific targets and tasks of different studies are often divergent, making direct comparison impossible.}} This issue is further compounded by the fact that many studies utilize self-collected open-source projects and applications rather than standardized benchmarks. \textbf{\emph{The widespread variation in tasks and targets,}} even when benchmarks are shared, underscores the diversity of program analysis tasks and highlights the need for further refinement and standardization of benchmark efforts in the field.
    
\vspace{3mm}
\begin{custommdframed}
\textbf{Finding 1 (Academic Program Analysis):} Research in program analysis is highly diverse, with studies often pursuing different objectives even when using the same benchmarks. While effectiveness is a consistent evaluation metric, some studies also assess time and memory efficiency.
\end{custommdframed}
\vspace{0.2cm}

\vspace{1mm}
\begin{custommdframed}
\textbf{Finding 2 (Academic Program Analysis):} Despite integration with intelligent technologies, the research paradigm remains rooted in traditional techniques like static and dynamic analysis. This innovation is also primarily applied to language-specific tasks, highlighting a significant gap in cross-language and general analysis research.
\end{custommdframed}
\vspace{1mm}

\subsection{Code Generation and Completion (61 papers, 4.5\%):}
\label{subsec:codeGen}

   \textbf{Unlike many other research fields, the evaluation of code generation and completion models relies heavily on established benchmarks}, which can be categorized into two groups according to their task scopes: stand-alone function-level and repository-level benchmarks.

\textbf{Stand-alone function-level benchmarks assess the generation of self-contained code snippets}. Representative examples include HumanEval~\cite{chen2021evaluating}, MBPP~\cite{austin2021program}, APPS~\cite{hendrycks2021measuring}, and CodeContests~\cite{li2022competition}. HumanEval evaluates functional correctness based on docstrings, a design further extended by HumanEval+~\cite{liu2023your} (with augmented test cases) and HumanEval-ET~\cite{dong2025codescore} (incorporating error-triggering scenarios). Similarly, MBPP (Mostly Basic Python Programs) and its variants test models on entry-level programming problems. For more complex challenges, APPS and CodeContests evaluate a model’s ability to solve more complex competitive programming problems, demanding advanced algorithmic reasoning.

\textbf{Repository-level benchmarks examine code completion in the context of larger software projects.} Commonly used benchmarks include RepoEval~\cite{zhang2023repocoder}, CrossCodeEval~\cite{ding2023crosscodeeval}, and M2RC-EVAL~\cite{liu-etal-2025-m2rc}. RepoEval evaluates both line-level and API-level completion, while CrossCodeEval specifically measures cross-file code completion performance. M2RC-EVAL is a multilingual repository-level benchmark covering 18 programming languages, with fine-grained annotations at both bucket-level and semantic-level across diverse completion scenarios.

We summarize several recent studies that use these overlapping benchmarks and report their evaluation results in Table~\ref{tab:functionLevelGen} and~\ref{tab:repoLevelGen}.
Table~\ref{tab:functionLevelGen} shows the performance of advanced approaches on five stand-alone function-level benchmarks. \textbf{\emph{Overall, no single approach achieves the best performance across all five benchmarks.}} Among them, the recent \texttt{$\mu$Fix}~\cite{tian2025fixing} (based on ChatGPT) performs best on HumanEval and its variants (reaching up to 90.24\% on the original HumanEval), while \texttt{PairCoder} achieves the top results on the MBPP series. \textbf{\emph{Notably, performance declines consistently on benchmarks marked with ``+'' or ``-ET'', which include more test cases and edge cases, indicating limited robustness in existing approaches.}}

For more complex competitive programming benchmarks, \texttt{$\mu$Fix} attains a \emph{pass@1} score of 35.67\% on a sampled subset of the \emph{APPS dataset} (300 problems) using ChatGPT, while \texttt{PairCoder}~\cite{zhang2024pair} achieves 17.95\% and 15.15\% on the \emph{CodeContests validation and test sets}, respectively, using GPT-3.5-turbo. \textbf{\emph{These results suggest that techniques for stand-alone function-level code generation are still far from mature, particularly in complex task settings such as the competitive tasks, and require further research.}}

Research on repository-level code completion has emerged more recently and remains relatively less explored than function-level generation. Table~\ref{tab:repoLevelGen} presents the performance of two advanced approaches—\texttt{RLCoder}~\cite{wang2024rlcoder} and \texttt{GraphCoder}~\cite{liu2024graphcoder}—on RepoEval and CrossCodeEval. Although these two approaches do not share exactly the same benchmarks, their results are still informative. \texttt{RLCoder} (based on DeepSeek-Coder-7B) achieves a higher Exact Match (EM) score on RepoEval line-level tasks (48.75) than on API-level tasks (39.88). In contrast, \texttt{GraphCoder} (using GPT-3.5-turbo) performs better on Java tasks, especially at the API-level (61.57 EM). On CrossCodeEval, \texttt{RLCoder} demonstrates moderate cross-file performance, with higher scores in Python (30.28 EM) than in Java (26.09 EM). \textbf{\emph{These outcomes underscore the ongoing challenges in project-level and cross-file code generation, revealing considerable room for improvement in exact match accuracy.}}
    
    \begin{table}[h]
        \centering
        \small
        \caption{Performance of Four Advanced Approaches on Five Standalone Function-Level Benchmarks (Pass@1\%)}
        \label{tab:functionLevelGen}
        \begin{tabular}{lcccc}
        \hline
        \textbf{Benchmark} & 
        \makecell{\textbf{FlowGen~\cite{lin2024soen}} \\ \textbf{(GPT-3.5-turbo)}} & 
        \makecell{\textbf{PairCoder~\cite{zhang2024pair}} \\ \textbf{(GPT-3.5-turbo)}} & 
        \makecell{\textbf{ClarifyGPT~\cite{mu2024clarifygpt}} \\ \textbf{(GPT-4)}} & 
        \makecell{\textbf{$\mu$Fix~\cite{tian2025fixing}} \\ \textbf{(ChatGPT)}} \\
        \hline
        HumanEval~\cite{chen2021evaluating} & 75.2\% & 87.80\% & 87.80\% & 90.24\% \\
        HumanEval+~\cite{liu2023your} & - & 77.44\% & - & 80.49\% \\
        HumanEval-ET~\cite{dong2025codescore} & 65.5\% & - & 78.05\% & 79.88\% \\
        \hline
        MBPP~\cite{austin2021program} & 82.5\% & 80.60\% & 78.69\% & - \\
        MBPP+~\cite{liu2023your} & - & 77.69\% & - & - \\
        MBPP-ET~\cite{dong2025codescore} & 56.7\% & - & 58.47\% & 69.10\% \\
        \hline
        \end{tabular}
    \end{table}
    
\begin{table}[h]
    \centering
    \caption{Performance of TWO Advanced Approaches on TWO Standalone Function-Level Benchmarks (EM / ES)}
    \label{tab:repoLevelGen}
    \begin{tabular}{lllcc}
    \hline
    \textbf{Benchmark} & \textbf{Task Type} & \textbf{Method/Model} & \textbf{EM} & \textbf{ES} \\
    \hline
    RepoEval-Original~\cite{zhang2023repocoder} & Line-level & RLCoder (DeepSeek-Coder-7B)~\cite{wang2024rlcoder} & 48.75 & 69.43 \\
    RepoEval-Original & API-level & RLCoder (DeepSeek-Coder-7B) & 39.88 & 66.22 \\
    RepoEval-Updated-Python & Line-level & GraphCoder (GPT-3.5-turbo)~\cite{liu2024graphcoder} & 46.60 & 69.42 \\
    RepoEval-Updated-Python & API-level & GraphCoder (GPT-3.5-turbo) & 45.25 & 66.81 \\
    RepoEval-Updated-Java & Line-level & GraphCoder (GPT-3.5-turbo) & 50.60 & 78.94 \\
    RepoEval-Updated-Java & API-level & GraphCoder (GPT-3.5-turbo) & 61.57 & 82.66 \\
    \hline
    CrossCodeEval-Python~\cite{ding2023crosscodeeval} & - & RLCoder (DeepSeek-Coder-7B) & 30.28 & 74.42 \\
    CrossCodeEval-Java & - & RLCoder (DeepSeek-Coder-7B) & 26.09 & 67.31 \\
    \hline
    \end{tabular}
\end{table}

\vspace{1mm}
\begin{custommdframed}
\textbf{Finding 1 (Academic Code Generation and Completion):} While emergent research in code synthesis has advanced repository-level code generation, techniques for stand-alone function-level generation remain immature. This is particularly evident in complex, competitive tasks such as CodeContests.
\end{custommdframed}
\vspace{0.2cm}

\vspace{1mm}
\begin{custommdframed}
\textbf{Finding 2 (Academic Code Generation and Completion):} The field predominantly employs functional correctness metrics, such as Pass@k and exact match (EM). However, there is a critical need to expand evaluation frameworks to include practical attributes like computational efficiency, security, and cost.
\end{custommdframed}
\vspace{1mm}

    \subsection{Automated Program Repair and Issue Resolution (126 papers, 9.2\%):} This category encompasses Automated Program Repair (APR) and security vulnerability detection \& fixing, as both aim to automatically identify and resolve source code issues. Research in these areas often utilizes shared benchmarks.

    \subsubsection{Common benchmarks and the corresponding approaches' capabilities}
    
    \textbf{For vulnerability detection, common datasets include Big-Vul ~\cite{fan2020ac} and Devign ~\cite{zhou2019devign}}. Big-Vul is a large-scale, imbalanced dataset comprising 188,636 C/C++ functions collected from real-world GitHub projects; it includes 10,900 (6\%) vulnerable and 177,736 (94\%) non-vulnerable functions, mirroring the real-world scarcity of vulnerable code. Devign is a balanced dataset consisting of 27,318 examples (53.22\% vulnerable) collected from two different large C programming-based projects: Qemu and FFmpeg. \textbf{\emph{The performance of different approaches on Big-Vul varies considerably.}} As shown in Table~\ref{tab:vul}, \texttt{DeepDFA} ~\cite{steenhoek2024dataflow} (using UniXcoder) achieves state-of-the-art results with an F1-score of 96.46\%, significantly outperforming other methods like \texttt{PDBERT} ~\cite{liu2024pre} (59.41\%), \texttt{SnoPY} ~\cite{cao2024snopy} (41.85\%), \texttt{CausalVul} ~\cite{rahman2024towards} (39\%) on the binary vulnerability classification task from Big-Vul. Besides, \texttt{MoEVD} ~\cite{yang2025one} achieves an F1-score of 44\% on the vulnerability type detection task from Big-Vul, which still leaves substantial room for improvement. \textbf{Research utilizing the Devign dataset is relatively limited.} Among the existing work, \texttt{PDBERT} and \texttt{CausalVul} demonstrate comparable performance, achieving F1-scores of 67.61\% and 68\% respectively. \textbf{\emph{The fact that approaches like \texttt{PDBERT} and \texttt{CausalVul} perform better on Devign than on Big-Vul reveals a performance gap. This indicates that vulnerability detection must be advanced to tackle the challenges of real-world, imbalanced datasets. Moreover, most approaches are limited to classifying \emph{whether a function is vulnerable} and do not provide explanations, limiting their usefulness for vulnerability fixing.}}  
    
    \textbf{For APR, popular benchmarks include QuixBugs ~\cite{lin2017quixbugs}, Defects4J ~\cite{just2014defects4j}, Vul4J ~\cite{bui2022vul4j}, and SWE-Bench ~\cite{jimenez2023swe}}. QuixBugs, consisting of 40 programs translated to both Python and Java, each with a bug on a single line. The bugs are common and basic fundamental errors, usually can be fixed within 1 minute ~\cite{lawler2012hire}. The Defects4J dataset comprises 835 real-world defects collected from 17 open-source Java projects. Specifically, Defects4J v1.2 includes 395 bugs from 6 projects, while Defects4J v2.0 expands the benchmark with an additional 440 bugs across 11 newly added projects, providing a broader and more diverse foundation for evaluating software testing and debugging techniques. Vul4J is a dataset of 79 real-world Java vulnerabilities from 51 projects, distinguished by including a Proof of Vulnerability (PoV) test case for each entry; and SWE-bench presents more complex, real-world software engineering problems drawn from open-source projects and provides a complete development context, enabling models to reason about multi-file dependencies, complex logic, and real-world project structures.

      \begin{table}[h]
    \centering
    \small
    \caption{Performance of Advanced \emph{Vulnerability Detection} Methods on the Benchmarks of Big-Vul and Devign}
    \label{tab:vul}
    \begin{tabular}{lccccc}
    \toprule
    \textbf{Benchmark} &\textbf{Method} & \textbf{Task}& \textbf{F1-Score} & \textbf{Precision} & \textbf{Recall} \\
    \midrule
    Big-Vul~\cite{fan2020ac} &SnoPY~\cite{cao2024snopy} &Binary Vulnerability Classification & 41.85 & 38.12 & 46.39 \\
    Big-Vul & PDBERT~\cite{liu2024pre} & Binary Vulnerability Classification & 59.41 & -&- \\
    Big-Vul & CausalVul~\cite{rahman2024towards} & Binary Vulnerability Classification & 39.00 & 43 & 36 \\
    Big-Vul & DeepDFA~\cite{steenhoek2024dataflow}  &Binary Vulnerability Classification& 96.46 & 97.82 & 95.14 \\
    Big-Vul & MoEVD\cite{yang2025one}  &Vulnerability Type Detection & 44.00 & 42 & 46 \\
    \hline
    Devign~\cite{zhou2019devign} & PDBERT  & Binary Vulnerability Classification &  67.61 & - &- \\
    Devign & CausalVul & Binary Vulnerability Classification & 68.00 & - &-\\
    \bottomrule
    \end{tabular}
    \end{table}


    The key results of advanced approaches on these APR benchmarks are summarized in Table ~\ref{tab:apr}. 
    \begin{itemize}[leftmargin=1.7em]
        \item On the simpler \emph{QuixBugs} dataset, LoRA-finetuned \texttt{CodeLlama-7B} achieves a repair rate of 65\% ~\cite{li2024exploring}, outperforming \texttt{GAMMA} ~\cite{zhang2023gamma} (55\%) while slightly below \texttt{TARE} ~\cite{zhu2023tare} (67.5\%).

        \item For the more challenging \emph{Defects4J} benchmark, \texttt{RepairAgent} ~\cite{bouzenia2024repairagent} attains a 19.6\% repair rate across versions 1.2 and 2.0 using GPT-3.5, supporting multi-line and multi-file fixes. \texttt{Neural Template Repair} ~\cite{huang2025template} leverages the large CodeLlama-70B model to achieve 35.2\% on Defects4J v1.2. The highest performance comes from \texttt{(IA)\textsuperscript{3}} ~\cite{li2024exploring}, which achieves 43.8\% on the single-hunk subset of Defects4J v2.0 using CodeLlama-7B, though limited to single-hunk repairs.

        \item On \emph{Vul4J}, which focuses specifically on security vulnerabilities, repair rates for single-hunk bugs are substantially lower: Neural Template Repair (using StarCoder-15B) leads at 31.4\%, followed by VulMaster ~\cite{zhou2024out} (25.7\%) and DeepVulGuard ~\cite{steenhoek2024closing} (13\%). \textbf{\emph{This demonstrates that vulnerability repair tasks are significantly more challenging than general program repair in Defects4J.}}

        \item   \textbf{\emph{The complex SWE-bench benchmark has attracted growing interest from both academia and industry \footnote{\url{https://www.swebench.com/}}}}, driving the development of improved methods to achieve higher performance. Among recent studies, \texttt{Agentless} ~\cite{xia2025demystifying} achieved a 50.8\% solve rate on the SWE-bench Verified set using Claude 3.5 Sonnet, while \texttt{SpecRover} ~\cite{ruan2024specrover} solved 19.31\% of the full SWE-bench set using a combination of Claude 3.5 Sonnet and GPT-4o.
    \end{itemize}

    \begin{table}[h]
    \centering
    \small
    \caption{Performance of Automated Program Repair Methods on Major Benchmarks.}
    \label{tab:apr}
    \begin{tabular}{lcccc}
    \toprule
    \textbf{Benchmark} & \textbf{Method} & \textbf{Resolved} & \textbf{Backbone Model} \\
    \midrule
    QuixBugs (40 bugs)~\cite{lin2017quixbugs} 
        & LoRA~\cite{li2024exploring} 
        & 65.0\% 
        & CodeLlama-7B \\
    QuixBugs (40 bugs) 
        & GAMMA~\cite{zhang2023gamma} 
        & 55.0\% 
        & GPT-J \\
    QuixBugs (40 bugs) 
        & TARE~\cite{zhu2023tare} 
        & 67.5\% 
        & CodeT5+ \\
    \midrule
    Defects4J v2.0 (217 single-hunk bugs)~\cite{just2014defects4j} & (IA)\textsuperscript{3}~\cite{li2024exploring} & 43.8\% & CodeLlama-7B \\
    Defects4J v1.2 (395 bugs) & Neural Template Repair~\cite{huang2025template} & 35.2\% & CodeLlama-70B \\
    Defects4J v1.2 + v2.0 (835 bugs) & RepairAgent~\cite{bouzenia2024repairagent} & 19.6\% & GPT-3.5 \\
    \midrule
    Vul4J (35 bugs)~\cite{bui2022vul4j} & Neural Template Repair & 31.4\% & StarCoder-15B \\
    Vul4J (35 bugs) & VulMaster~\cite{zhou2024out} & 25.7\% & CodeT5p-large \\
    Vul4J (24 bugs) & DeepVulGuard~\cite{steenhoek2024closing} & 13.0\% & CodeBERT \\
    \midrule
    SWE-bench Verified (500 tasks)~\cite{jimenez2023swe} & Agentless~\cite{xia2025demystifying} & 50.8\% & Claude 3.5 Sonnet \\
    SWE-bench (2294 tasks) & SpecRover~\cite{ruan2024specrover} & 19.31\% & Claude 3.5 Sonnet + GPT-4o \\
    \bottomrule
    \end{tabular}
    \end{table}

\vspace{3mm}
\begin{custommdframed}
\textbf{Finding 1 (Academic Automated Program Repair and Issue Resolution):} The evolution of benchmarks from Defects4J to SWE-bench shows a clear trend toward increasingly complex and realistic issues. However, the issue resolution rate remains low, as evidenced by scores of 19.31\% ~\cite{ruan2024specrover} and 33.83\% (the top leaderboard entry as of 2025-10-23) on the full SWE-bench, indicating a pressing need for significant improvement, particularly for multi-hunk and multi-file bugs.
\end{custommdframed}
\vspace{0.2cm}

\vspace{1mm}
\begin{custommdframed}
\textbf{Finding 2 (Academic Automated Program Repair and Issue Resolution):} Research in the field predominantly focuses on defect detection and repair rates, leaving substantial gaps in the interpretability and cost-related aspects of automated program repair. Furthermore, existing benchmarks are predominantly limited to C/C++, Java, and Python, with a notable lack of support for other low-resource programming languages (e.g., ArkTS) and cross-language scenarios.
\end{custommdframed}
\vspace{5mm}
    
    \subsection{Software Dependency Management (41 papers, 3.0\%):}
    \label{subsec: Dependency}
    Recent research in version and dependency management has advanced across several key areas, including \emph{dependency analysis and optimization}, \emph{dependency migration},  \emph{dependency compatibility and conflict resolution}, \emph{vulnerability and security risk mitigation}, \emph{software build and reproducibility}, and \emph{semantic version compliance}. However, \textbf{\emph{this field is characterized by a lack of standardized benchmarks.}} Studies typically construct their own evaluation datasets, likely due to the contextual nature of dependency issues and the difficulty in obtaining large-scale, real-world ground-truth data. This reliance on custom benchmarks complicates the direct comparison of different approaches and challenges the generalizability of reported results.

    \begin{itemize}[leftmargin=1.7em]
        \item In \textbf{dependency analysis and optimization}, studies such as \texttt{Drosos's work} ~\cite{drosos2024bloat} and \texttt{Slimming}~\cite{song2024efficiently} have revealed \textbf{\emph{severe dependency bloat and redundancy across ecosystems}}. The former, analyzing 1,302 Python projects, found that 51\% of dependencies were redundant and that many security vulnerabilities resided in unused code, leading to practical optimization results accepted by developers (30 of 36 PRs merged). \texttt{Slimming} enhanced \textbf{\emph{dependency trimming}} through reflection analysis, achieving 97.0\% precision and 98.8\% recall on a large-scale Java benchmark.

        \item Research on \textbf{dependency migration} focuses on ensuring safe and efficient transitions to newer or alternative libraries. The work of Mujahid et al. ~\cite{mujahid2023go} proposed a data-driven approach to identify declining npm packages and recommend reliable substitutes with 96\% accuracy, supported by developer validation, while \texttt{GoblinUpdater} ~\cite{jaime2024balancing} optimized update quality-cost trade-offs in 107 Maven projects, improving dependency freshness by up to two orders of magnitude without sacrificing build stability. 
         
        \item  The field of \textbf{dependency compatibility and conflict resolution} has been advanced by automated techniques such as \texttt{LooCo} ~\cite{wang2023automatically} and \texttt{HERA} ~\cite{xie2024pet}. \texttt{LooCo} addresses conflicts \textbf{\emph{within a single repository}}; evaluated on benchmarks of 83 conflict issues and 104 prone-to-misconfiguration libraries, it resolved 54.8\% of conflicts, surpassing baseline methods by 21 issues on average. In contrast, \texttt{HERA} specifically targets \textbf{\emph{cross-repository compatibility}}. On a benchmark of 1,692 real-world cases, it demonstrated high precision (90.5\%) and recall (93.7\%), and successfully fixed 26 of 27 real-world incidents collected from GitHub and Stack Overflow.

        \item In \textbf{vulnerability and security management}, \texttt{Ranger}~\cite{zhang2023mitigating} \textbf{\emph{mitigated open-source vulnerability}} persistence in the Maven ecosystem by restoring 75.64\% of version ranges and automatically patching over 90\% of affected projects, while studies such as He et al. ~\cite{he2025pinning} and Hu et al. ~\cite{hu2024empirical} revealed that \textbf{\emph{naive version pinning}} can increase security risk and maintenance cost, and that distributed ecosystems such as Go suffer from slower vulnerability propagation and repair due to publication and indexing delays. 
    
        \item The topic of \textbf{software build and reproducibility} emphasizes deterministic and auditable builds. \texttt{CXXCrafter}~\cite{yu2025cxxcrafter}, an LLM-based agent for C/C++ project building, achieved a 78\% build success rate across 752 open-source projects, far exceeding GPT-4o and default build commands. Complementarily, \texttt{Benedetti's study} ~\cite{benedetti2025empirical} across six major ecosystems demonstrated that \textbf{\emph{reproducibility varies significantly}}, with Cargo and npm near 100\% and others initially below 12\%, yet showed that \textbf{\emph{minor toolchain adjustments can raise reproducibility above 90\%.}}

        \item Finally, \textbf{change impact analysis and semantic version compliance} has been examined through large-scale empirical studies. \texttt{GoSVI}~\cite{li2023large}, based on 124,000 Go libraries and 532,000 clients, achieved 91\% accuracy in detecting breaking changes and revealed that 28.6\% of non-major upgrades violate semantic versioning, affecting 33.3\% of clients. Similarly, \texttt{Jayasuriya's research} ~\cite{jayasuriya2024understanding} in Java ecosystems (8,086 Maven projects) showed that 2.3\% of dependency updates introduce behavioral regressions, two-thirds of which occur in non-major updates, indicating \textbf{\emph{persistent violations of semantic versioning principles}}. Together, these studies collectively establish a comprehensive landscape of dependency management challenges and advances across ecosystems, emphasizing \textbf{\emph{the critical interplay between automation, safety, and maintainability in modern software supply chains.}}
        
    \end{itemize}


\vspace{3mm}
\begin{custommdframed}
\textbf{Finding (Academic Software Dependency Management)}: A key challenge is the lack of standardized benchmarks, complicating the comparison and generalizability of results, alongside persistent issues such as dependency bloat, violations of semantic versioning, slow vulnerability resolution, and security risks introduced by common practices like version pinning.
\end{custommdframed}
\vspace{1mm}

\subsection{Pre-trained Code Models (48 papers, 3.5\%):} \label{subsec:pretrainedModels}
    Pre-trained code models serve as the foundation for various code intelligence tasks, including code search, completion, synthesis, and summarization. Within the realm of SE4AI (Software Engineering for AI), a critical research direction is \textbf{enhancing the robustness and efficiency of these models. This focus on the model's foundational capabilities sets it apart from tasks like code synthesis and summarization, which operate at the application level and are evaluated on their output.}
    
    Pre-trained code models have been extensively evaluated across a series of well-established benchmarks designed to assess various program understanding and generation capabilities. The major benchmark categories include: (1) \textbf{Clone Detection}, such as BigCloneBench ~\cite{svajlenko2014towards} (a large-scale dataset for detecting functionally similar code clones) and POJ-104 (the clone detection task provided by CodeXGLUE ~\cite{lu2021codexglue}); (2) \textbf{Code Search}, typically evaluated on CodeSearchNet ~\cite{husain2019codesearchnet} (a large collection of code functions paired with natural language documentation), including its challenging variants AdvTest ~\cite{lu2021codexglue} and cosQA ~\cite{huang2021cosqa}; (3) \textbf{Code Translation}, mainly assessed on the Java–C\# dataset from CodeXGLUE; and (4) \textbf{Code Summarization}, also based on CodeSearchNet, evaluating natural language summarization of code snippets. These benchmarks together provide a comprehensive evaluation framework for different downstream tasks in code intelligence.     
    Table~\ref{tab:pretrainbench} summarizes the performance of several recent methods across these benchmarks.   

    \begin{itemize}[leftmargin=1.7em]
    \item For \textbf{clone detection}, \texttt{Tailor}~\cite{liu2023learning} achieves outstanding performance on BigCloneBench, reaching a precision of 99.72\%, a recall of 99.9\%, and an F1-score of 99.8\%, demonstrating near-perfect detection capability. The \texttt{Adapter} approach based on GraphCodeBERT~\cite{liu2023empirical} also performs strongly on the same benchmark, achieving a precision of 95.32\%, a recall of 95.30\%, and an F1-score of 95.30\%. 
    
    On POJ-104, \texttt{REPEAT}~\cite{gao2023keeping} (built upon CodeT5) achieves a precision of 91.05\%, a recall of 91.77\%, and an F1-score of 91.26\%, indicating robust clone identification capability. \texttt{ContraBERT}~\cite{liu2023contrabert} further enhances clone retrieval performance by improving the CodeBERT baseline’s MAP@R from 84.29\% to 90.46\%, and the GraphCodeBERT baseline’s MAP@R from 85.16\% to 90.06\%. \texttt{TRACED}~\cite{ding2024traced} achieves a MAP@R of 91.2\% on the same dataset, reflecting their strong retrieval effectiveness. \textbf{\emph{However, as evidenced by Krinke et al. ~\cite{krinke2025misuse}, these near-perfect results may be inflated by dataset-specific biases and the flawed ground truth of the benchmarks, rather than solely reflecting genuine semantic understanding.}} 
    
    \item In the domain of \textbf{code search}, adapter-based fine-tuning of \texttt{UniXcoder} has been shown to outperform full-parameter tuning ~\cite{wang2023one}, increasing MRR from 74.5\% to 75.3\% on the CodeSearchNet benchmark. Further improvements are achieved by \texttt{GrammarT5} ~\cite{zhu2024grammart5}, particularly on the challenging AdvTest and cosQA datasets, where it raises MRR from 41.30\% to 44.12\% and from 70.10\% to 73.48\%, respectively. It should be noted that these results are limited to pre-trained code models from the three top conferences in recent years and their reported common-used benchmarks like CodeSearchNet. Consequently, \textbf{\emph{while incremental progress is evident, the performance of code search techniques remains imperfect—a limitation that may adversely impact downstream tasks which depend on it, such as those within Retrieval-Augmented Generation (RAG) frameworks for code generation and repair.}}
    
    \item In \textbf{code translation tasks}, \texttt{ContraBERT} ~\cite{liu2023contrabert} improves the baseline of GraphCodeBERT on BLEU from 80.58 to 80.78 and accuracy from 59.4 to 59.9 on the Java→C\# task from CodeXGLUE, while on the C\#→Java task, it improves \texttt{GraphCodeBERT} baseline's BLEU from 72.64 to 76.24 and accuracy from 58.80 to 60.50. Evaluated on the same benchmark, \texttt{GrammarT5} ~\cite{zhu2024grammart5} achieves higher BLEU, with 91.31 BLEU for Java→C\#, and 90.53 BLEU for C\#→Java. \textbf{\emph{As complex, repository-level code translation benchmarks like RepoTransBench ~\cite{wang2024repotransbenchrealworldbenchmarkrepositorylevel} emerge, we observe a pressing need for more relevant tasks that accurately reflect these realistic and complex scenarios.}}
    
    \item For \textbf{code summarization}, \texttt{adapter-based fine-tuning} provides consistent gains ~\cite{wang2023one}, improving CodeT5's BLEU score from 20.23 to 20.42 on the Java summarization task and from 20.26 to 20.52 on the Python summarization task, both from CodeSearchNet. \texttt{GrammarT5} reaches BLEU scores of 20.66 on Java and 20.21 on Python, while \texttt{REPEAT} (CodeT5)~\cite{gao2023keeping} further enhances performance to 24.79 and 21.99, respectively.
    \textbf{\emph{Despite recent advances in methods such as REPEAT and GrammarT5 have led to certain improvements in BLEU scores, the overall progress remains limited. The code summarization task still holds significant potential for further enhancement.}}
    \end{itemize}
    
    \vspace{0.2em}
    Besides the above code-direct downstream tasks, \textbf{the training and deployment} of the code models also raise critical concerns regarding \textbf{model memorization and privacy risks}. Several research focus on this aspect. \texttt{Nie's work} ~\cite{nie2024decoding} explores how code large language models (Code LLMs) memorize and reproduce sensitive information such as API keys. By analyzing token-level features, the study successfully differentiates real secrets from model-generated fabrications. \texttt{Chen et al.} ~\cite{chen2024promise} assess collaborative training methods (e.g., centralized, federated, and incremental), using cross-organizational code data. The study reveals federated learning maintains competitive accuracy with reduced memorization, unlike centralized training. It highlights the critical trade-off between model utility and privacy. \texttt{Yang et al.} ~\cite{yang2024unveiling} empirically analyze memorization in pre-trained code models. They categorize memorized content into 14 types, identify influencing factors like model size and output length, and assess inference metrics. Results show models can reproduce licensed, vulnerable, or sensitive code, underscoring the need for deduplication and careful preprocessing.
    
    Collectively, these studies reveal that \textbf{memorization in code models is a widespread and critical issue, leading to the leakage of sensitive information, licensed code, and vulnerable snippets}. Existing research has begun to develop detection methods, mitigation strategies, such as data deduplication ~\cite{yang2024unveiling}, federated learning ~\cite{chen2024promise}, and token-level guided decoding ~\cite{nie2024decoding}, and ethical guidelines ~\cite{nie2024decoding} to address these risks. \textbf{\emph{However, the persistent tension between model utility and privacy preservation underscores the need for continued efforts in designing privacy-aware training frameworks and robust evaluation mechanisms}}.

    \begin{table}[h]
    \centering
    \small
    \caption{Performance of Several Code Pre-training Models on Major Downstream Tasks}
    \label{tab:pretrainbench}
    \begin{tabular}{p{3cm}lll}
    \toprule
    \textbf{Benchmark} & \textbf{Method} & \textbf{Task} & \textbf{Main Results} \\
    \midrule
    BigCloneBench~\cite{svajlenko2014towards} & Tailor~\cite{liu2023learning} & Clone Detection & P:99.72/R:99.9/F1:99.8 \\
    BigCloneBench & Adapter (GraphCodeBERT)~\cite{liu2023empirical} & Clone Detection & P:95.32/R:95.30/F1:95.30 \\ 
    POJ-104~\cite{lu2021codexglue} & REPEAT (CodeT5)~\cite{gao2023keeping} & Clone Detection & P:91.05/R:91.77/F1:91.26 \\
    POJ-104 & ContraBERT~\cite{liu2023contrabert} / TRACED~\cite{ding2024traced} & Clone Detection & MAP@R $\approx$ 91 \\ \hline
    CodeSearchNet~\cite{husain2019codesearchnet} & Adapter (UniXcoder)~\cite{wang2023one} & Code Search & MRR: 74.5 $\rightarrow$ 75.3 \\
    CodeSearchNet (Adv~\cite{lu2021codexglue}+cosQA~\cite{huang2021cosqa}) & GrammarT5~\cite{zhu2024grammart5} & Code Search & MRR +2.8$\sim$3.4↑ \\ \hline
    CodeXGLUE (Java$\leftrightarrow$C\#)~\cite{lu2021codexglue} & GrammarT5 & Code Translation & BLEU:90.5–91.3 \\
    CodeXGLUE (Java$\leftrightarrow$C\#) & ContraBERT & Code Translation & BLEU:76.2–80.8, Acc:59.9-60.5 \\ \hline
    CodeSearchNet & Adapter~\cite{wang2023one} / GrammarT5 / REPEAT & Code Summarization & BLEU:20.2–24.8 \\
    \bottomrule
    \end{tabular}
    \end{table}
    
\vspace{3mm}
\begin{custommdframed}
\textbf{Finding (Academic Pre-trained Code Models):} While emerging techniques show impressive improvements on widely-used benchmarks, this progress underscores the need for more complex and influential challenges, such as repository-level translation.
Moreover, these common benchmarks focus almost exclusively on evaluating functional correctness, paying little attention to equally important concerns like intellectual property and private code protection. 
\end{custommdframed}
\vspace{1mm}

\section{Survey Feedback from Industry}
\label{sec:survey}

To gather practical feedback and identify industry needs regarding the six prominent research topics identified in Section \ref{sec:academicAnalysis}, we designed a dedicated questionnaire for each topic. Each questionnaire addressed three core aspects: \ding{172} the current adoption of relevant tools and technologies (including specific tasks they are applied to), \ding{173} identified needs for improvement (across dimensions such as efficiency, cost, and security), and \ding{174} the perceived industrial relevance of the academic topic. All questions were presented in single-choice or multiple-choice format, with an optional open-ended ``Other'' field for additional suggestions. The complete questionnaires are provided in our replication package.

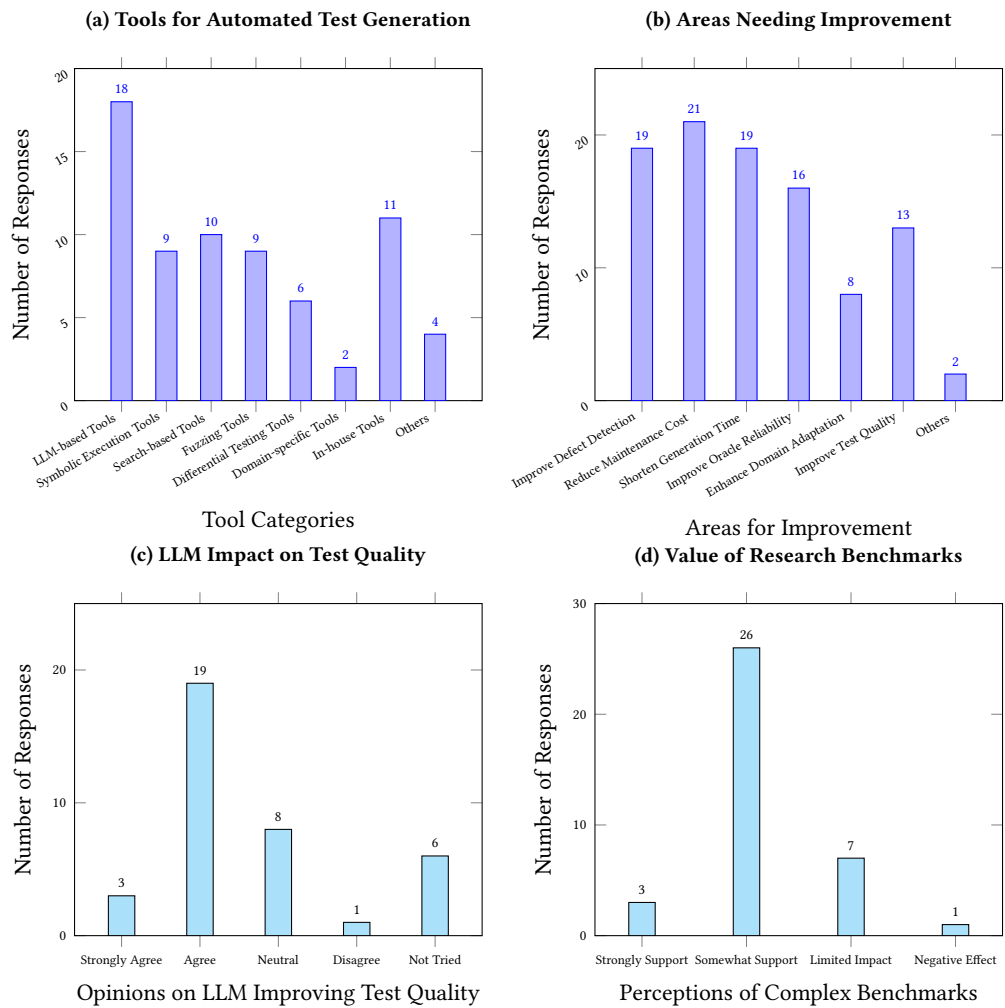
\begin{figure}[!htbp]
    \centering
    \begin{tikzpicture}
    
    \begin{axis}[
        scale=1,
        name=plot1,
        at={(0,0)},
        ybar,
        ymin=0,
        ymax=20,
        bar width=8pt,
        width=7cm,
        height=6cm,
        symbolic x coords={
            LLM-based Tools,
            Symbolic Execution Tools,
            Search-based Tools,
            Fuzzing Tools,
            Differential Testing Tools,
            Domain-specific Tools,
            In-house Tools,
            Others
        },
        xtick=data,
        nodes near coords,
        nodes near coords align={vertical},
        enlarge x limits=0.15,
        ylabel={Number of Responses},
        xlabel={Tool Categories},
        tick label style={font=\tiny, rotate=30, anchor=east, align=center},
        title={(a) Tools for Automated Test Generation},
        title style={font=\small\bfseries, yshift=5pt},
        every node near coord/.append style={font=\tiny},
    ]
    \addplot coordinates {
        (LLM-based Tools,18)
        (Symbolic Execution Tools,9)
        (Search-based Tools,10)
        (Fuzzing Tools,9)
        (Differential Testing Tools,6)
        (Domain-specific Tools,2)
        (In-house Tools,11)
        (Others,4)
    };
    \end{axis}

    \begin{axis}[
        scale=1,
        name=plot2,
        at={(plot1.east)},
        anchor=west,
        xshift=1.5cm,
        ybar,
        ymin=0,
        ymax=25,
        bar width=8pt,
        width=7cm,
        height=6cm,
        symbolic x coords={
            Improve Defect Detection,
            Reduce Maintenance Cost,
            Shorten Generation Time,
            Improve Oracle Reliability,
            Enhance Domain Adaptation,
            Improve Test Quality,
            Others
        },
        xtick=data,
        nodes near coords,
        nodes near coords align={vertical},
        enlarge x limits=0.15,
        ylabel={Number of Responses},
        xlabel={Areas for Improvement},
        tick label style={font=\tiny, rotate=30, anchor=east, align=center},
        title={(b) Areas Needing Improvement},
        title style={font=\small\bfseries, yshift=5pt},
        every node near coord/.append style={font=\tiny},
    ]
    \addplot coordinates {
        (Improve Defect Detection,19)
        (Reduce Maintenance Cost,21)
        (Shorten Generation Time,19)
        (Improve Oracle Reliability,16)
        (Enhance Domain Adaptation,8)
        (Improve Test Quality,13)
        (Others,2)
    };
    \end{axis}

    \begin{axis}[
        scale=1,
        name=plot3,
        at={(plot1.south)},
        anchor=north,
        yshift=-2.7cm,
        ybar,
        ymin=0,
        ymax=25,
        bar width=10pt,
        width=7cm,
        height=6cm,
        symbolic x coords={
            Strongly Agree,
            Agree,
            Neutral,
            Disagree,
            Not Tried
        },
        xtick=data,
        nodes near coords,
        nodes near coords align={vertical},
        enlarge x limits=0.15,
        ylabel={Number of Responses},
        xlabel={Opinions on LLM Improving Test Quality},
        tick label style={font=\tiny, rotate=0},
        title={(c) LLM Impact on Test Quality},
        title style={font=\small\bfseries, yshift=5pt},
        every node near coord/.append style={font=\tiny},
    ]
    \addplot[fill=cyan!30] coordinates {
        (Strongly Agree,3)
        (Agree,19)
        (Neutral,8)
        (Disagree,1)
        (Not Tried,6)
    };
    \end{axis}

    \begin{axis}[
        scale=1,
        name=plot4,
        at={(plot3.east)},
        anchor=west,
        xshift=1.5cm,
        ybar,
        ymin=0,
        ymax=30,
        bar width=10pt,
        width=7cm,
        height=6cm,
        symbolic x coords={
            Strongly Support,
            Somewhat Support,
            Limited Impact,
            Negative Effect
        },
        xtick=data,
        nodes near coords,
        nodes near coords align={vertical},
        enlarge x limits=0.15,
        ylabel={Number of Responses},
        xlabel={Perceptions of Complex Benchmarks},
        tick label style={font=\tiny, rotate=0},
        title={(d) Value of Research Benchmarks},
        title style={font=\small\bfseries, yshift=5pt},
        every node near coord/.append style={font=\tiny},
    ]
    \addplot[fill=cyan!30] coordinates {
        (Strongly Support,3)
        (Somewhat Support,26)
        (Limited Impact,7)
        (Negative Effect,1)
    };
    \end{axis}

    \end{tikzpicture}
    \caption{Survey Results on Automated Test Generation Tools and Practices (n=37)}
    \label{fig:survey-results-test}
\end{figure}

We received 282 responses from 17 industrial organizations, including Intel, Microsoft, Alibaba, Huawei, Tencent, ByteDance, and Meituan, as well as several state key system engineering institutions of China. To ensure the validity of the feedback, we used a targeted sampling approach: participants were selected from our professional network based on their routine tasks and relevant experience, and each respondent received only the questionnaires pertinent to their field. For example, the questionnaire on testing was sent exclusively to participants who had experience with testing-related tasks.
The results for each topic are presented in the following sections, and the partial results for each topic are presented in Figures~\ref{fig:survey-results-test}--\ref{fig:dependency-management-survey}.

\subsection{Automated Testing  Technology (37 Responses)}
Among all participants, 23 participants (62.16\%) reported that their teams utilize automated testing techniques, while the remainder continue to rely on manual testing. This finding suggests that automated testing technologies are still far from maturity, indicating substantial market potential for high-quality solutions in this domain. Whether for manual or automated testing, software requirements are the most common input for automated testing (26 responses, 70.27\%), while about half of the respondents also frequently use source code, API specifications, and existing test cases.

    \subsubsection{Feedback on the Application of Automated Tools.} 
    As shown in Fig.~\ref{fig:survey-results-test} (a), the techniques used in industrial are diverse, including LLM-based (18, 48.65\%), symbolic execution-based (9, 24.32\%), and search-based approaches (10, 27.03\%). Besides, domain-specific techniques like ArchiDroid for Android are used too (2, 5.41\%). In addition, 11 respondents (29.73\%) reported a preference for using internally developed tools, which are neither open-source nor commercial but are instead custom-built to align closely with specific business needs.

    Practitioners expected several key desired improvements on automated testing techniques (Fig.~\ref{fig:survey-results-test} (b)): reducing testing costs (56.76\%), improving defect detection while reducing false positives (51.35\%), enhancing test case quality (43.24\%), and strengthening domain adaptation capabilities (21.62\%).
     
     
     \subsubsection{Feedback on Academic Focus.} Some research claims that ``\emph{LLMs can significantly enhance the compilation pass rate and readability of generated test cases}'' ~\cite{jiang2024towards,rao2023cat}. \textbf{\emph{Practitioners' agreement is nuanced; while nearly half generally acknowledge some improvement, they adopt a more pragmatic and guarded stance.}} For instance, 19 participants agree with the claim in principle, noting that \emph{while the compilation pass rate improves, the test cases still require significant manual justification.} Eight participants remain neutral, perceiving no significant quality difference between LLM-generated and traditional test cases.

    Another academic trend involves \emph{the use of complex benchmarks, such as real-world, large-scale applications}~\cite{wang2025llmdroid,liu2023fill}. Only three participants highly appraise this direction. While the majority (26 participants, 70.27\%) agree that it could help align academic research with industrial problems, \textbf{\emph{they caution that `realistic' academic systems still differ from complex industrial environments. More seriously, seven participants believe the impact is limited}}, stating that research driven by the need to showcase technical novelty for publications often yields solutions that do not scale in practical settings.

   \vspace{3mm}
    \begin{custommdframed}
    \textbf{Findings on automated testing technology in industry:} This study concludes with three key findings:
    (1) Specialists widely use automated tools, with a 62.16\% adoption rate confirming a robust market; 
    (2) The use of LLMs in testing is still nascent, requiring enhancements in both effectiveness and efficiency; 
    (3) Practitioners remain skeptical of academic claims, citing a disconnect between technical novelty and practical applicability in complex real-world settings.
    \end{custommdframed}
    \vspace{0cm}

    \subsection{Program Analysis (40 Responses).}

\begin{figure}[!htbp]
    \centering
    \begin{tikzpicture}

    \begin{axis}[
        scale=1,
        name=plot1,
        at={(0,0)},
        ybar,
        ymin=0, ymax=40,
        bar width=6pt,
        width=5.5cm, height=4.5cm,
        symbolic x coords={Static Analysis, Dynamic Analysis, Hybrid Analysis},
        xtick=data,
        nodes near coords,
        enlarge x limits=0.12,
        ylabel={Number of Responses},
        xlabel={Common Analysis Types},
        tick label style={font=\tiny, rotate=25, anchor=east},
        title={(a) Analysis Types},
        title style={font=\small\bfseries, yshift=5pt},
        every node near coord/.append style={font=\tiny},
    ]
    \addplot coordinates {(Static Analysis,33)(Dynamic Analysis,15)(Hybrid Analysis,11)};
    \end{axis}

    \begin{axis}[
        scale=1,
        name=plot2,
        at={(plot1.east)},
        anchor=west,
        xshift=1.2cm,
        ybar,
        ymin=0, ymax=40,
        bar width=6pt,
        width=5.5cm, height=4.5cm,
        symbolic x coords={Traditional Only, Traditional + AI/LLM, AI/LLM Only, Uncertain},
        xtick=data,
        nodes near coords,
        enlarge x limits=0.12,
        ylabel={Number of Responses},
        xlabel={Usage of Analysis Techniques},
        tick label style={font=\tiny, rotate=25, anchor=east},
        title={(b) Analysis Techniques},
        title style={font=\small\bfseries, yshift=5pt},
        every node near coord/.append style={font=\tiny},
    ]
    \addplot coordinates {(Traditional Only,15)(Traditional + AI/LLM,17)(AI/LLM Only,4)(Uncertain,4)};
    \end{axis}

    \begin{axis}[
        scale=1,
        name=plot3,
        at={(plot2.east)},
        anchor=west,
        xshift=1.2cm,
        ybar,
        ymin=0, ymax=30,
        bar width=8pt,
        width=5.5cm, height=4.5cm,
        symbolic x coords={
            Core Analysis Capability,
            Integration and Usability,
            Intelligent Support,
            Performance and Reliability,
            Adaptability
        },
        xtick=data,
        nodes near coords,
        nodes near coords align={vertical},
        enlarge x limits=0.12,
        ylabel={Number of Responses},
        xlabel={Desired Enhancements},
        tick label style={font=\tiny, rotate=25, anchor=east},
        title={(c) Tool Enhancements},
        title style={font=\small\bfseries, yshift=5pt},
        every node near coord/.append style={font=\tiny},
    ]
    \addplot coordinates {
        (Core Analysis Capability,28)
        (Integration and Usability,18)
        (Intelligent Support,22)
        (Performance and Reliability,13)
        (Adaptability,6)
    };
    \end{axis}

    \begin{axis}[
        scale=1,
        name=plot4,
        at={(plot1.south)},
        anchor=north,
        yshift=-2.8cm,
        ybar,
        ymin=0, ymax=25,
        bar width=8pt,
        width=5.5cm, height=4.5cm,
        symbolic x coords={Highly Reflective, Partially Reflective, Rarely Reflective, Uncertain},
        xtick=data,
        nodes near coords,
        enlarge x limits=0.12,
        ylabel={Number of Responses},
        xlabel={Research Claims Reflection},
        tick label style={font=\tiny, rotate=25, anchor=east},
        title={(d) Academic Claims},
        title style={font=\small\bfseries, yshift=5pt},
        every node near coord/.append style={font=\tiny},
    ]
    \addplot[fill=cyan!30] coordinates {(Highly Reflective,3)(Partially Reflective,23)(Rarely Reflective,7)(Uncertain,7)};
    \end{axis}

    \begin{axis}[
        scale=1,
        name=plot5,
        at={(plot4.east)},
        anchor=west,
        xshift=1.2cm,
        ybar,
        ymin=0, ymax=30,
        bar width=8pt,
        width=5.5cm, height=4.5cm,
        symbolic x coords={
            Highly Reliable,
            Potential but Risky,
            Useful for Assistance,
            Poor Explainability
        },
        xtick=data,
        nodes near coords,
        enlarge x limits=0.12,
        ylabel={Number of Responses},
        xlabel={LLM Reliability \& Explainability},
        tick label style={font=\tiny, rotate=25, anchor=east},
        title={(e) LLM Reliability},
        title style={font=\small\bfseries, yshift=5pt},
        every node near coord/.append style={font=\tiny},
    ]
    \addplot[fill=cyan!30] coordinates {
        (Highly Reliable,3)
        (Potential but Risky,24)
        (Useful for Assistance,11)
        (Poor Explainability,2)
    };
    \end{axis}

    \begin{axis}[
        scale=1,
        name=plot6,
        at={(plot5.east)},
        anchor=west,
        xshift=1.2cm,
        ybar,
        ymin=0, ymax=12,
        bar width=8pt,
        width=5.5cm, height=4.5cm,
        symbolic x coords={
            Technical,
            Integration,
            Trust,
            Resource,
            Research-Industry Gap
        },
        xtick=data,
        nodes near coords,
        enlarge x limits=0.12,
        ylabel={Number of Responses},
        xlabel={Barriers to Adoption},
        tick label style={font=\tiny, rotate=25, anchor=east},
        title={(f) Adoption Barriers},
        title style={font=\small\bfseries, yshift=5pt},
        every node near coord/.append style={font=\tiny},
    ]
    \addplot[fill=cyan!30] coordinates {
        (Technical,10)
        (Integration,9)
        (Trust,8)
        (Resource,8)
        (Research-Industry Gap,5)
    };
    \end{axis}

    \end{tikzpicture}
    \caption{Survey Results on Program Analysis Practices and Perceptions (n=40)}
    \label{fig:combined-analysis-survey}
\end{figure}
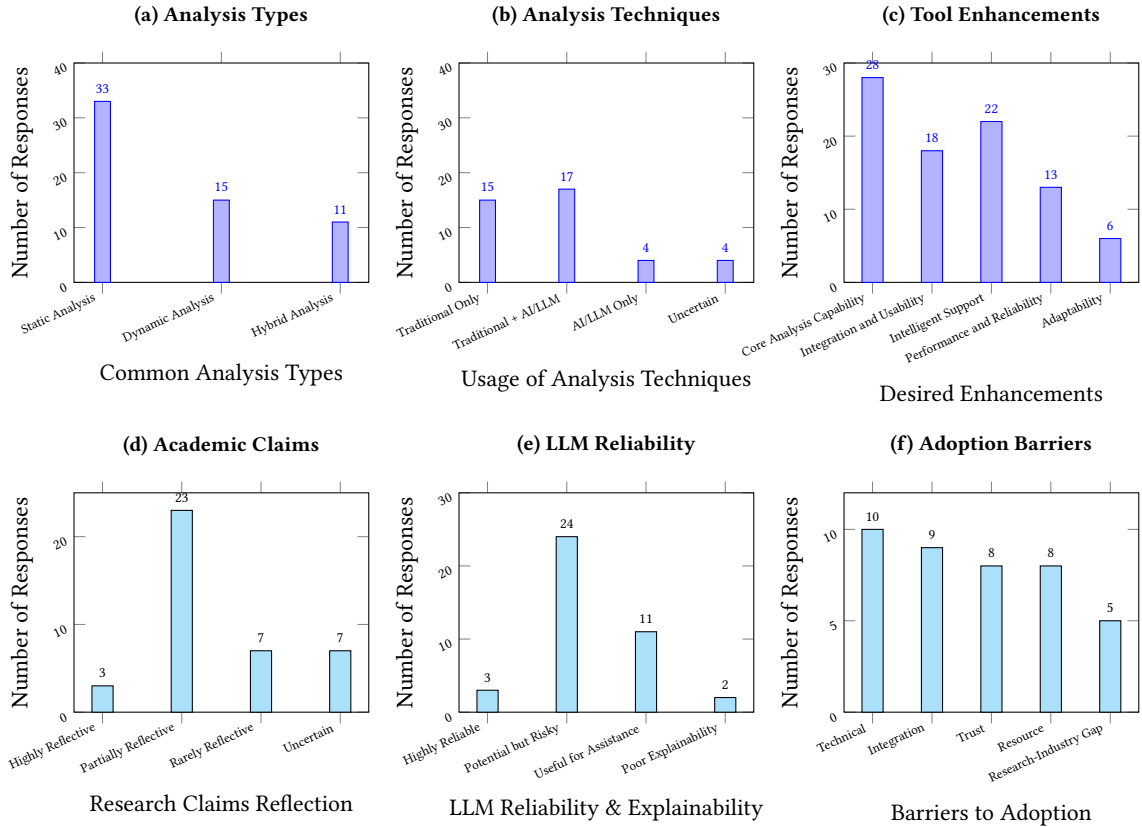

   The primary results from industry feedback on program analysis are shown in Fig.~\ref{fig:combined-analysis-survey}. Subfigures (a)-(c) pertain to the use of automated tools in daily work, while (d)-(f) address three key academic research areas.
    
    \subsubsection{Feedback on the Application of Automated Tools.} The vast majority of participant groups (90.0\%) employ automated program analysis techniques. As seen from Fig.~\ref{fig:combined-analysis-survey} (a), static analysis (e.g., code scanning) is the most prevalent (82.5\%), followed by dynamic analysis (e.g., runtime monitoring, 37.5\%) and hybrid methods that combine both (27.5\%). In terms of technical approaches (Fig.~\ref{fig:combined-analysis-survey} (b)), traditional techniques (e.g., data flow and syntax tree analysis) remain primary, used by 80\% of respondents (32 participants). However, a significant portion (42.5\%, 17 participants) reported incorporating AI/LLM-augmented techniques, with four groups indicating they rely entirely on LLM-based methods. \textbf{\emph{This trend converges with the prevailing academic research paradigm, which seeks to enhance traditional analysis through intelligent techniques.}}

        Regarding desired improvements (Fig.~\ref{fig:combined-analysis-survey} (c)), participants highlighted several key features: enhanced core analysis capabilities (e.g., inter-procedural and cross-language analysis, 70.0\%), smarter features (e.g., automated fixes, 55.0\%), better integration and user experience (e.g., IDE integration and visualization, 45.0\%), improved performance and reliability (e.g., higher speed and fewer false positives, 32.5\%), and greater adaptability (e.g., support for more languages and frameworks, 12.5\%). \textbf{\emph{Collectively, the feedback underscores a practitioner focus that balances technical power with practical efficiency: advanced core analysis is essential, but its value is fully realized only when coupled with intelligent automation and frictionless integration into the developer's workflow.}}

        \subsubsection{Feedback on Academic Focus.}  We obtained industry perspectives on three academic focuses. 
        
        First, \emph{several recent studies claimed that their tools or approaches achieve ``high precision'' or even ``zero false positives'' on specific datasets.} For instance, Optional Checker reached 93\% precision in detecting Java Optional misuse ~\cite{yoo2024verifying}, and WeMinT ~\cite{meng2023wemint} demonstrated a precision of 100\% in detecting App Secret leaks in WeChat Mini-Programs, with zero false positives among the reported cases. We inquired to what extent practitioners believe these claims reflect real-world industrial performance. As shown in Fig.~\ref{fig:combined-analysis-survey}(d), overall, \textbf{\emph{the respondents expressed skepticism.}} Specifically, 56\% indicated that the results only partially reflect practical performance, noting that real-world factors such as code scale, external dependencies, and specific configurations significantly hinder the effectiveness of academic approaches. A further 17.5\% believed the results are not applicable in practice, citing a substantial gap between academic benchmarks and industrial scenarios.

        Secondly, given the growing use of LLMs to augment traditional program analysis, \emph{we sought industry attitudes toward LLM-augmented approaches.} The results are shown in subfigure (e). Only three participants (7.5\%) expressed high reliability in these methods, stating that \emph{the semantic understanding capability of LLMs can fundamentally address issues that are difficult to model with traditional analysis.} Meanwhile, \textbf{\emph{the majority (60\%) recognized some potential but emphasized that the ``hallucination'' problem necessitates strict verification of outputs using traditional methods}}, casting doubt on their reliability. 27.5\% considered LLMs more suitable for auxiliary roles at present, noting they are more applicable to supportive, heuristic tasks (e.g., generating summaries or candidates) rather than serving as core analytical logic. Finally, 5\% remained skeptical, stating that the ``black-box'' nature of LLMs fundamentally conflicts with the ``explainability'' required in program analysis, making large-scale application difficult. 
    
        Finally, practitioners also \emph{ranked obstacles to adopting academic prototypes in large-scale industrial applications} (seen in subfigure (f)). The key barriers include: \textbf{technical limitations} (efficiency/accuracy/performance not meeting industrial needs, 25\%), \textbf{integration challenges} (difficulty integrating with existing tools and workflows, 22.5\%), \textbf{cognitive bias} (deep-seated distrust due to high false positive rates, 20\%), \textbf{resource barriers} (high computational and memory costs, 20\%), and \textbf{problem-solving misalignment} (academic research not addressing critical industrial pain points, 12.5\%). 

         \vspace{3mm}
        \begin{custommdframed}
        \textbf{Findings on program analysis in industry:}  
        (1) There is high industry adoption (90\%) of automated program analysis, with a clear trend toward AI/LLM-augmented traditional techniques. 
        (2) Although advanced core capabilities are paramount, the ultimate utility of these tools hinges on user-centric features like automated fixes and workflow integration.
        (3) A persistent gap exists between academic progress and industrial acceptance, as practitioners remain cautious due to the complexities of real-world applications.
        \end{custommdframed}
        \vspace{0cm}

    \subsection{Code Generation and Completion (52 Responses).} 

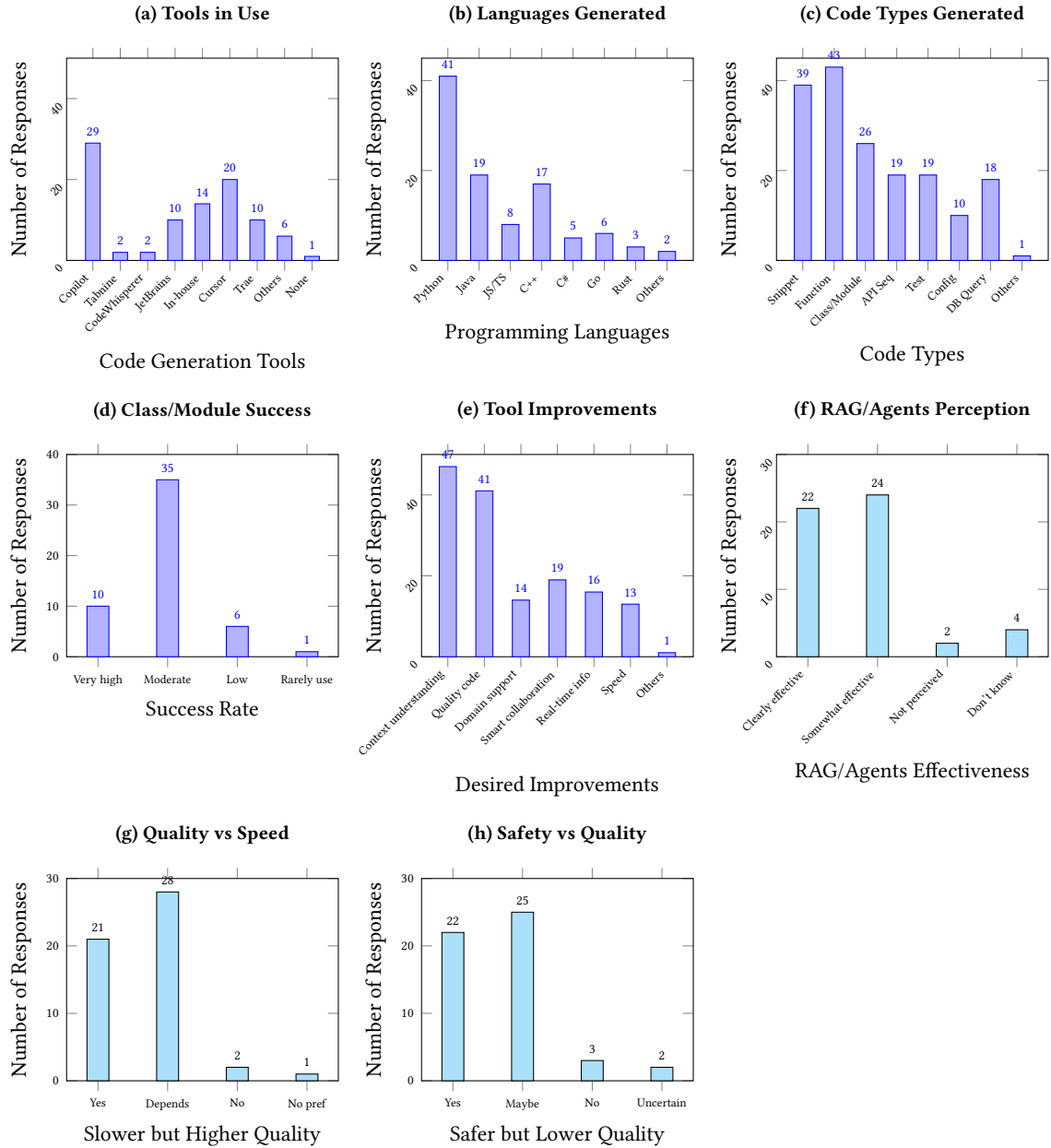
\begin{figure}[H]
    \centering
    \begin{tikzpicture}
    
    \begin{axis}[
        scale=1,
        name=plot1,
        at={(0,0)},
        ybar,
        ymin=0,
        ymax=50,
        bar width=6pt,
        width=5.5cm,
        height=4.5cm,
        symbolic x coords={Copilot, Tabnine, CodeWhisperer, JetBrains, In-house, Cursor, Trae, Others, None},
        xtick=data,
        nodes near coords,
        nodes near coords align={vertical},
        enlarge x limits=0.12,
        ylabel={Number of Responses},
        xlabel={Code Generation Tools},
        tick label style={font=\tiny, rotate=45, anchor=east},
        title={(a) Tools in Use},
        title style={font=\small\bfseries, yshift=5pt},
        every node near coord/.append style={font=\tiny},
    ]
    \addplot coordinates {(Copilot,29)(Tabnine,2)(CodeWhisperer,2)(JetBrains,10)(In-house,14)(Cursor,20)(Trae,10)(Others,6)(None,1)};
    \end{axis}

    \begin{axis}[
        scale=1,
        name=plot2,
        at={(plot1.east)},
        anchor=west,
        xshift=1.2cm,
        ybar,
        ymin=0,
        ymax=45,
        bar width=7pt,
        width=5.5cm,
        height=4.5cm,
        symbolic x coords={Python, Java, JS/TS, C++, C\#, Go, Rust, Others},
        xtick=data,
        nodes near coords,
        nodes near coords align={vertical},
        enlarge x limits=0.12,
        ylabel={Number of Responses},
        xlabel={Programming Languages},
        tick label style={font=\tiny, rotate=45, anchor=east},
        title={(b) Languages Generated},
        title style={font=\small\bfseries, yshift=5pt},
        every node near coord/.append style={font=\tiny},
    ]
    \addplot coordinates {(Python,41)(Java,19)(JS/TS,8)(C++,17)(C\#,5)(Go,6)(Rust,3)(Others,2)};
    \end{axis}

    \begin{axis}[
        scale=1,
        name=plot3,
        at={(plot2.east)},
        anchor=west,
        xshift=1.2cm,
        ybar,
        ymin=0,
        ymax=45,
        bar width=7pt,
        width=5.5cm,
        height=4.5cm,
        symbolic x coords={Snippet, Function, Class/Module, API Seq, Test, Config, DB Query, Others},
        xtick=data,
        nodes near coords,
        nodes near coords align={vertical},
        enlarge x limits=0.12,
        ylabel={Number of Responses},
        xlabel={Code Types},
        tick label style={font=\tiny, rotate=45, anchor=east},
        title={(c) Code Types Generated},
        title style={font=\small\bfseries, yshift=5pt},
        every node near coord/.append style={font=\tiny},
    ]
    \addplot coordinates {(Snippet,39)(Function,43)(Class/Module,26)(API Seq,19)(Test,19)(Config,10)(DB Query,18)(Others,1)};
    \end{axis}

    \begin{axis}[
        scale=1,
        name=plot4,
        at={(plot1.south)},
        anchor=north,
        yshift=-2.8cm,
        ybar,
        ymin=0,
        ymax=40,
        bar width=9pt,
        width=5.5cm,
        height=4.5cm,
        symbolic x coords={Very high, Moderate, Low, Rarely use},
        xtick=data,
        nodes near coords,
        nodes near coords align={vertical},
        enlarge x limits=0.15,
        ylabel={Number of Responses},
        xlabel={Success Rate},
        tick label style={font=\tiny, rotate=0},
        title={(d) Class/Module Success},
        title style={font=\small\bfseries, yshift=5pt},
        every node near coord/.append style={font=\tiny},
    ]
    \addplot coordinates {(Very high,10)(Moderate,35)(Low,6)(Rarely use,1)};
    \end{axis}

    \begin{axis}[
        scale=1,
        name=plot5,
        at={(plot4.east)},
        anchor=west,
        xshift=1.2cm,
        ybar,
        ymin=0,
        ymax=50,
        bar width=7pt,
        width=5.5cm,
        height=4.5cm,
        symbolic x coords={Context understanding, Quality code, Domain support, Smart collaboration, Real-time info, Speed, Others},
        xtick=data,
        nodes near coords,
        nodes near coords align={vertical},
        enlarge x limits=0.12,
        ylabel={Number of Responses},
        xlabel={Desired Improvements},
        tick label style={font=\tiny, rotate=45, anchor=east},
        title={(e) Tool Improvements},
        title style={font=\small\bfseries, yshift=5pt},
        every node near coord/.append style={font=\tiny},
    ]
    \addplot coordinates {(Context understanding,47)(Quality code,41)(Domain support,14)(Smart collaboration,19)(Real-time info,16)(Speed,13)(Others,1)};
    \end{axis}

    \begin{axis}[
        scale=1,
        name=plot6,
        at={(plot5.east)},
        anchor=west,
        xshift=1.2cm,
        ybar,
        ymin=0,
        ymax=30,
        bar width=9pt,
        width=5.5cm,
        height=4.5cm,
        symbolic x coords={Clearly effective, Somewhat effective, Not perceived, Don't know},
        xtick=data,
        nodes near coords,
        nodes near coords align={vertical},
        enlarge x limits=0.15,
        ylabel={Number of Responses},
        xlabel={RAG/Agents Effectiveness},
        tick label style={font=\tiny, rotate=45, anchor=east, align=center},
        title={(f) RAG/Agents Perception},
        title style={font=\small\bfseries, yshift=5pt},
        every node near coord/.append style={font=\tiny},
    ]
    \addplot[fill=cyan!30] coordinates {(Clearly effective,22)(Somewhat effective,24)(Not perceived,2)(Don't know,4)};
    \end{axis}

    \begin{axis}[
        scale=1,
        name=plot7,
        at={(plot4.south)},
        anchor=north,
        yshift=-3.2cm,
        ybar,
        ymin=0,
        ymax=30,
        bar width=9pt,
        width=5.5cm,
        height=4.5cm,
        symbolic x coords={Yes, Depends, No, No pref},
        xtick=data,
        nodes near coords,
        nodes near coords align={vertical},
        enlarge x limits=0.15,
        ylabel={Number of Responses},
        xlabel={Slower but Higher Quality},
        tick label style={font=\tiny, rotate=0},
        title={(g) Quality vs Speed},
        title style={font=\small\bfseries, yshift=5pt},
        every node near coord/.append style={font=\tiny},
    ]
    \addplot[fill=cyan!30] coordinates {(Yes,21)(Depends,28)(No,2)(No pref,1)};
    \end{axis}

    \begin{axis}[
        scale=1,
        name=plot8,
        at={(plot7.east)},
        anchor=west,
        xshift=1.2cm,
        ybar,
        ymin=0,
        ymax=30,
        bar width=9pt,
        width=5.5cm,
        height=4.5cm,
        symbolic x coords={Yes, Maybe, No, Uncertain},
        xtick=data,
        nodes near coords,
        nodes near coords align={vertical},
        enlarge x limits=0.15,
        ylabel={Number of Responses},
        xlabel={Safer but Lower Quality},
        tick label style={font=\tiny, rotate=0},
        title={(h) Safety vs Quality},
        title style={font=\small\bfseries, yshift=5pt},
        every node near coord/.append style={font=\tiny},
    ]
    \addplot[fill=cyan!30] coordinates {(Yes,22)(Maybe,25)(No,3)(Uncertain,2)};
    \end{axis}


    \end{tikzpicture}
    \caption{Survey Results on Code Generation Tools Usage and Perceptions (n=52)}
    \label{fig:combined-codegen-survey}
\end{figure}

    Figure~\ref{fig:combined-codegen-survey} presents industrial feedback on code generation and completion. Subfigures (a)-(e) detail the usage of automated tools, while (f)-(h) evaluate three key academic focuses.
    
 \subsubsection{Feedback on the Application of Automated Tools.} Based on the survey responses, \textbf{\emph{GitHub Copilot is the most widely used tool for code generation and completion (55.77\%), followed by Cursor (38.46\%) and various internally developed tools (26.92\%) (Fig.~\ref{fig:combined-codegen-survey}(a))}}. In addition to popular academic languages such as Python, Java, and C++, participants also reported generating JavaScript/TypeScript (15.38\%), Go (11.54\%), C\# (9.6\%), Rust (5.77\%), Lua (1.92\%), and Swift (1.92\%) (Fig.~\ref{fig:combined-codegen-survey} (b)). The types of code generated are diverse too, including stand-alone functions (82.69\%), single lines or statements (75\%), classes or modules (50\%), API call sequences (36.54\%), test cases (36.54\%), database queries such as SQL (34.62\%), and configuration files (19.23\%) (Fig.~\ref{fig:combined-codegen-survey} (c)).

        Regarding the experience of using tools to generate class or module code, 19.23\% of participants gave highly positive feedback, while \textbf{\emph{the majority (78.85\%) found the success rate to be moderate, typically requiring modifications before use}} (Fig.~\ref{fig:combined-codegen-survey} (d)). As shown in Fig. ~\ref{fig:combined-codegen-survey} (e), participants expect improvements in the following areas: better interpretation of project context (e.g., cross-file retrieval, 90.38\%), generation of high-quality code with strong performance, security, and maintainability (78.85\%), intelligent iteration based on feedback (36.54\%), integration of up-to-date ecosystem information (e.g., avoiding obsolete APIs, 30.77\%), support for specific domains such as cloud computing (26.92\%), and faster response times (25\%).

        \subsubsection{Feedback on Academic Focus.} Industrial perspectives were solicited on three prominent academic research themes.
        
        The first theme concerned the perception of \emph{retrieval-augmented generation (RAG) and multi-agent collaboration}, a hot topic in code generation tasks \cite{yu2024droidcoder,zhang2024pair,zhang2024instruct}. As shown in the subfigure (f), \textbf{\emph{a majority of practitioners acknowledged the role of these techniques;}} 22 participants (42.31\%) reported a clear perception of their effectiveness, whereas 46.15\% recognized their presence but deemed their impact limited. 
        
        The second theme concerned \emph{the trade-off between response efficiency and generated code quality}, a topic that has been the focus of multiple studies~\cite{wei2023towards,sun2024neural}. As shown in the subfigure (g), \textbf{\emph{the prevailing view was that this balance is highly context-dependent}}, with tolerance for slower responses in tasks requiring careful scrutiny but not in rapid development. Two respondents indicated a preference for manual coding over slow automated responses.
        
      The final theme addressed \emph{the trade-off involving security, reliability, and quality}, an area of growing recent concern~\cite{sun2023codemark,spiess2024calibration}. When asked about accepting code with slightly lower quality for enhanced security and reliability, \textbf{\emph{a pragmatic, conditional acceptance was observed}}: 42.30\% unconditionally prioritized code quality, and 48.08\% were accepting provided the quality degradation remained within a tolerable threshold (seen in subfigure (h)).

    \vspace{3mm}
    \begin{custommdframed}
    \textbf{Findings on code generation and completion in industry:}     
    (1) A strong preference exists for tools that are deeply integrated into the development workflow and native IDE environment, with GitHub Copilot and Cursor serving as prime examples.     
    (2) The practical evaluation of generated code is primarily based on the manual effort required for modification, driving an urgent need for models that achieve higher quality through enhanced project context interpretation and version-compatible API knowledge. 
    (3) While practitioners acknowledge the value of techniques like RAG and multi-agent systems, their acceptance of the associated performance trade-offs is strictly conditional; slower response times are only tolerable for non-time-sensitive tasks and within a strict threshold.
   \end{custommdframed}
    \vspace{0cm}

\subsection{Automated Program Repair and Issue Resolution (58 Responses).} 
\label{subsec:programRepair}

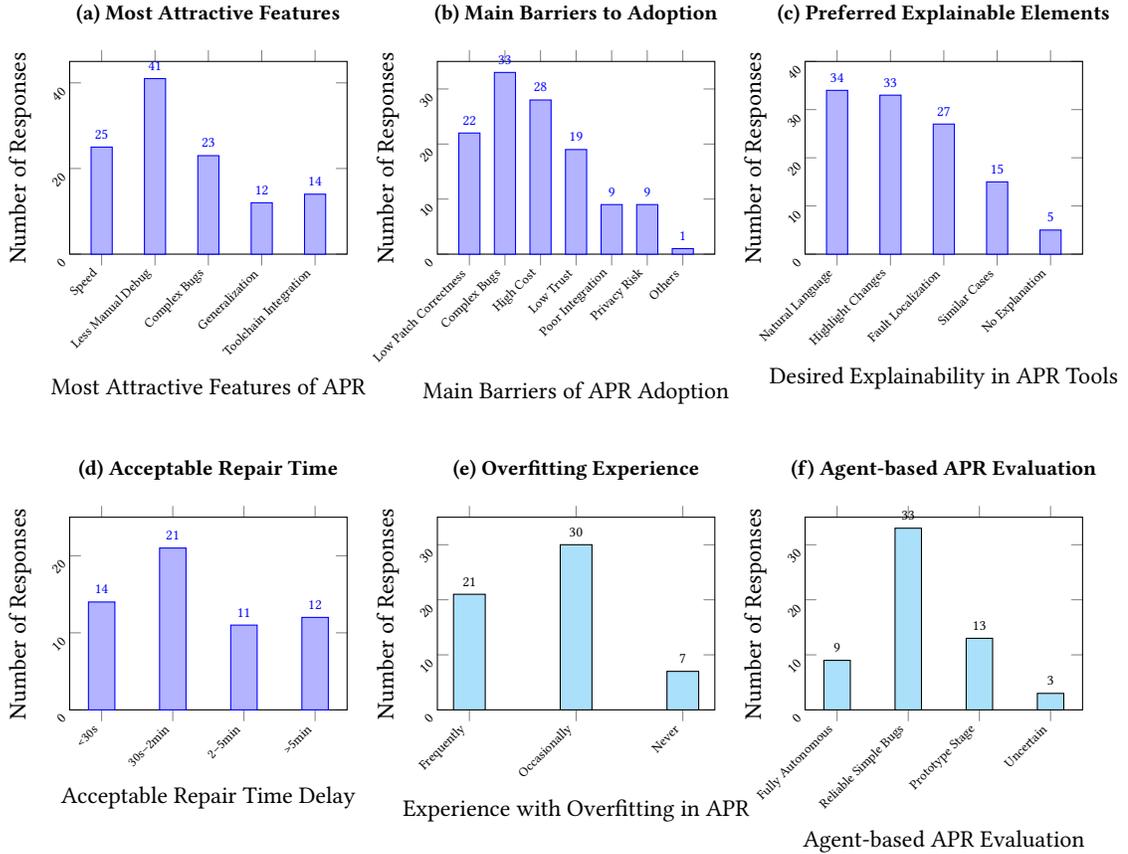
\begin{figure}[!htbp]
    \centering
    \begin{tikzpicture}
    
    \begin{axis}[
        scale=0.75,
        name=plot1,
        at={(0,0)},
        ybar,
        ymin=0,
        ymax=45,
        bar width=8pt,
        width=6.5cm,
        height=5cm,
        symbolic x coords={Speed, Less Manual Debug, Complex Bugs, Generalization, Toolchain Integration},
        xtick=data,
        nodes near coords,
        nodes near coords align={vertical},
        enlarge x limits=0.15,
        ylabel={Number of Responses},
        xlabel={Most Attractive Features of APR},
        tick label style={font=\tiny, rotate=45, anchor=east, align=center},
        title={(a) Most Attractive Features},
        title style={font=\small\bfseries, yshift=5pt},
        every node near coord/.append style={font=\tiny},
    ]
    \addplot coordinates {(Speed,25)(Less Manual Debug,41)(Complex Bugs,23)(Generalization,12)(Toolchain Integration,14)};
    \end{axis}

    \begin{axis}[
        scale=0.75,
        name=plot2,
        at={(plot1.east)},
        anchor=west,
        xshift=1.2cm,
        ybar,
        ymin=0,
        ymax=35,
        bar width=8pt,
        width=6.5cm,
        height=5cm,
        symbolic x coords={Low Patch Correctness, Complex Bugs, High Cost, Low Trust, Poor Integration, Privacy Risk, Others},
        xtick=data,
        nodes near coords,
        nodes near coords align={vertical},
        enlarge x limits=0.15,
        ylabel={Number of Responses},
        xlabel={Main Barriers of APR Adoption},
        tick label style={font=\tiny, rotate=45, anchor=east, align=center},
        title={(b) Main Barriers to Adoption},
        title style={font=\small\bfseries, yshift=5pt},
        every node near coord/.append style={font=\tiny},
    ]
    \addplot coordinates {(Low Patch Correctness,22)(Complex Bugs,33)(High Cost,28)(Low Trust,19)(Poor Integration,9)(Privacy Risk,9)(Others,1)};
    \end{axis}

    \begin{axis}[
        scale=0.75,
        name=plot3,
        at={(plot2.east)},
        anchor=west,
        xshift=1.2cm,
        ybar,
        ymin=0,
        ymax=40,
        bar width=8pt,
        width=6.5cm,
        height=5cm,
        symbolic x coords={Natural Language, Highlight Changes, Fault Localization, Similar Cases, No Explanation},
        xtick=data,
        nodes near coords,
        nodes near coords align={vertical},
        enlarge x limits=0.15,
        ylabel={Number of Responses},
        xlabel={Desired Explainability in APR Tools},
        tick label style={font=\tiny, rotate=45, anchor=east, align=center},
        title={(c) Preferred Explainable Elements},
        title style={font=\small\bfseries, yshift=5pt},
        every node near coord/.append style={font=\tiny},
    ]
    \addplot coordinates {(Natural Language,34)(Highlight Changes,33)(Fault Localization,27)(Similar Cases,15)(No Explanation,5)};
    \end{axis}

    \begin{axis}[
        scale=0.75,
        name=plot4,
        at={(plot1.south)},
        anchor=north,
        yshift=-3.5cm,  
        ybar,
        ymin=0,
        ymax=25,
        bar width=10pt,
        width=6.5cm,
        height=5cm,
        symbolic x coords={<30s, 30s–2min, 2–5min, >5min},
        xtick=data,
        nodes near coords,
        nodes near coords align={vertical},
        enlarge x limits=0.15,
        ylabel={Number of Responses},
        xlabel={Acceptable Repair Time Delay},
        tick label style={font=\tiny, rotate=45, anchor=east, align=center},
        title={(d) Acceptable Repair Time},
        title style={font=\small\bfseries, yshift=5pt},
        every node near coord/.append style={font=\tiny},
    ]
    \addplot coordinates {(<30s,14)(30s–2min,21)(2–5min,11)(>5min,12)};
    \end{axis}

    \begin{axis}[
        scale=0.75,
        name=plot5,
        at={(plot4.east)},
        anchor=west,
        xshift=1.2cm,
        ybar,
        ymin=0,
        ymax=35,
        bar width=12pt,
        width=6.5cm,
        height=5cm,
        symbolic x coords={Frequently, Occasionally, Never},
        xtick=data,
        nodes near coords,
        nodes near coords align={vertical},
        enlarge x limits=0.15,
        ylabel={Number of Responses},
        xlabel={Experience with Overfitting in APR},
        tick label style={font=\tiny, rotate=45, anchor=east, align=center},
        title={(e) Overfitting Experience},
        title style={font=\small\bfseries, yshift=5pt},
        every node near coord/.append style={font=\tiny},
    ]
    \addplot[fill=cyan!30] coordinates {(Frequently,21)(Occasionally,30)(Never,7)};
    \end{axis}

    \begin{axis}[
        scale=0.75,
        name=plot6,
        at={(plot5.east)},
        anchor=west,
        xshift=1.2cm,
        ybar,
        ymin=0,
        ymax=35,
        bar width=10pt,
        width=6.5cm,
        height=5cm,
        symbolic x coords={Fully Autonomous, Reliable Simple Bugs, Prototype Stage, Uncertain},
        xtick=data,
        nodes near coords,
        nodes near coords align={vertical},
        enlarge x limits=0.15,
        ylabel={Number of Responses},
        xlabel={Agent-based APR Evaluation},
        tick label style={font=\tiny, rotate=45, anchor=east, align=center},
        title={(f) Agent-based APR Evaluation},
        title style={font=\small\bfseries, yshift=5pt},
        every node near coord/.append style={font=\tiny},
    ]
    \addplot[fill=cyan!30] coordinates {(Fully Autonomous,9)(Reliable Simple Bugs,33)(Prototype Stage,13)(Uncertain,3)};
    \end{axis}

    \end{tikzpicture}
    \caption{Survey Results on Automated Program Repair (APR) Tools and Practices (n=58)}
    \label{fig:combined-apr-survey-results}
\end{figure}

Fig.~\ref{fig:combined-apr-survey-results} shows feedback from industrial participants on automated issue resolution. Subfigures (a) to (d), highlighted in light purple, present findings on the usage of automated tools, while the remaining subfigures, (e) and (f), highlighted in light blue, focus on two academic topics. 

    \subsubsection{Feedback on the Application of Automated Tools.} Survey results indicate \emph{a strong adoption of Automated Program Repair (APR) tools}, with 72.41\% of participants using them in their groups compared to 27.59\% who patch manually. The principal advantages of APR tools, as perceived by respondents (subfigure (a)), are the reduction of human effort (70.69\%), increased repair speed (43.10\%), and the capacity to resolve complex vulnerabilities (39.66\%). Secondary benefits include integration with existing toolchains (24.14\%) and cross-project generality (20.69\%). \textbf{\emph{These user priorities should guide the evaluation of academic APR approaches, emphasizing these practical benefits.}}


     However, \emph{a series of obstacles} hinder the adoption of current APR techniques by practitioners. As shown in the subfigure (b), the primary obstacles are \textbf{\emph{the inability to repair complex, multi-location issues (56.90\%), high computational cost (48.28\%), and insufficient patch accuracy (37.93\%)}}. Furthermore, a significant portion of respondents cited \textbf{\emph{a lack of developer trust}}, stemming from the "black-box" nature and poor explainability of the tools (32.76\%), alongside concerns over data privacy and code leakage (15.52\%). One participant elaborated that advanced techniques often require LLMs to process entire corporate repositories. This poses a dual challenge: it is technically difficult for the model to interpret such a large context, and it raises confidentiality concerns, as internal code cannot be shared with public LLM APIs.

    To address this trust deficit, we investigated the explanatory information essential for a generated patch (shown in subfigure (c)). The most requested elements were: the \textbf{\emph{repair rationale (58.62\%), highlighted code changes with context (56.90\%), auxiliary information on the issue's location (46.55\%), and historical similar cases (25.86\%)}}.

    Given the ongoing research into \emph{accelerating LLM-based approaches}~\cite{xia2025demystifying,ruan2024specrover}, we also gauged practitioners' performance expectations (shown in subfigure (d)). A plurality of respondents (36.21\%) deemed a latency of 30 seconds to 2 minutes acceptable. The responses were distributed across shorter (under 30 seconds, 24.14\%) and longer thresholds (2 to 5 minutes, 18.97\%; over 5 minutes, 20.69\%). \textbf{\emph{These findings establish a critical benchmark for optimizing future APR systems.}}

    \subsubsection{Feedback on Academic Focus.} Our survey gathered practitioner perspectives on two prominent academic focus. First, we investigated the significance of \emph{patch overfitting}, where a patch passes verification tests but fails to correct the root cause or introduces new faults~\cite{yang2023large,parasaram2023rete}. As shown in the subfigure (e), a combined 87.93\% of respondents encounter this problem, with a majority (51.72\%) describing it as a frequent issue requiring an urgent solution. \textbf{\emph{This strong consensus underscores the high practical value and necessity of research aimed at mitigating overfitting.}}

    Second, we solicited feedback on the industrial readiness of so-called \emph{efficient and human-like repair agents}~\cite{bouzenia2024repairagent,ruan2024specrover}. As shown in the subfigure (f), the prevailing view (56.90\%) is that their performance is currently confined to simple, common vulnerabilities, while a significant 22.41\% consider them mere toy demonstrations, not yet viable for real-world application. \textbf{\emph{These results send a clear message: for these techniques to transition from research to practice, future work must prioritize robustness and generalization to complex, real-world scenarios.}}

\vspace{3mm}
\begin{custommdframed}
\textbf{Findings on automated program repair and issue resolution in industry:}  
(1) Technical Capabilities: APR tools must evolve to effectively repair multi-location issues, reduce computational overhead, enhance patch accuracy, and foster developer trust. 
(2) Adoption Barriers: Critical challenges impeding routine use are the trust deficit, high costs, poor integration into development workflows, and data security concerns. 
(3) Research Direction: Future research on advanced techniques, such as repair agents, must prioritize evaluation in complex, real-world environments to demonstrate practical utility.
\end{custommdframed}
\vspace{0cm}


    \subsection{Software Dependency Management (58 Responses).} 

\begin{figure}[!htbp]
    \centering
    \begin{tikzpicture}
    
    \begin{axis}[
        scale=0.78,
        name=plot1,
        at={(0,0)},
        ybar,
        ymin=0,
        ymax=35,
        bar width=8pt,
        width=6.5cm,
        height=5cm,
        symbolic x coords={Security Vulnerabilities, Breaking Changes, Conflicts, Dependency Bloat, Non-reproducible Builds},
        xtick=data,
        nodes near coords,
        nodes near coords align={vertical},
        enlarge x limits=0.15,
        ylabel={Number of Responses},
        xlabel={Dependency Management Issues},
        tick label style={font=\tiny, rotate=45, anchor=east, align=center},
        title={(a) Frequent Dependency Issues},
        title style={font=\small\bfseries, yshift=5pt},
        every node near coord/.append style={font=\tiny},
    ]
    \addplot coordinates {(Security Vulnerabilities,29)(Breaking Changes,29)(Conflicts,32)(Dependency Bloat,21)(Non-reproducible Builds,7)};
    \end{axis}

    \begin{axis}[
        scale=0.78,
        name=plot2,
        at={(plot1.east)},
        anchor=west,
        xshift=1cm,
        ybar,
        ymin=0,
        ymax=40,
        bar width=8pt,
        width=6.5cm,
        height=5cm,
        symbolic x coords={Auto Vulnerability Detection, Recommend Updates, Remove Unused, Reproducible Builds, Cross-ecosystem},
        xtick=data,
        nodes near coords,
        nodes near coords align={vertical},
        enlarge x limits=0.15,
        ylabel={Number of Responses},
        xlabel={Desired Tool Features},
        tick label style={font=\tiny, rotate=45, anchor=east, align=center},
        title={(b) Key Desired Features},
        title style={font=\small\bfseries, yshift=5pt},
        every node near coord/.append style={font=\tiny},
    ]
    \addplot coordinates {(Auto Vulnerability Detection,37)(Recommend Updates,36)(Remove Unused,34)(Reproducible Builds,23)(Cross-ecosystem,16)};
    \end{axis}

    \begin{axis}[
        scale=0.78,
        name=plot3,
        at={(plot2.east)},
        anchor=west,
        xshift=1cm,
        ybar,
        ymin=0,
        ymax=30,
        bar width=10pt,
        width=6.5cm,
        height=5cm,
        symbolic x coords={Frequently, Occasionally, Rarely, Never, Unknown},
        xtick=data,
        nodes near coords,
        nodes near coords align={vertical},
        enlarge x limits=0.15,
        ylabel={Number of Responses},
        xlabel={SemVer Trust Paradox Experience},
        tick label style={font=\tiny, rotate=45, anchor=east, align=center},
        title={(c) SemVer Trust Paradox},
        title style={font=\small\bfseries, yshift=5pt},
        every node near coord/.append style={font=\tiny},
    ]
    \addplot[fill=cyan!30] coordinates {(Frequently,12)(Occasionally,26)(Rarely,12)(Never,2)(Unknown,6)};
    \end{axis}

    \begin{axis}[
        scale=0.78,
        name=plot4,
        at={(plot1.south)},
        anchor=north,
        yshift=-3.5cm,
        ybar,
        ymin=0,
        ymax=30,
        bar width=10pt,
        width=6.5cm,
        height=5cm,
        symbolic x coords={Severe, Acceptable, No time/effort yet, None, Unknown},
        xtick=data,
        nodes near coords,
        nodes near coords align={vertical},
        enlarge x limits=0.15,
        ylabel={Number of Responses},
        xlabel={Perceived Severity of Dependency Bloat},
        tick label style={font=\tiny, rotate=45, anchor=east, align=center},
        title={(d) Dependency Bloat Severity},
        title style={font=\small\bfseries, yshift=5pt},
        every node near coord/.append style={font=\tiny},
    ]
    \addplot[fill=cyan!30] coordinates {(Severe,11)(Acceptable,24)(No time/effort yet,15)(None,3)(Unknown,5)};
    \end{axis}

    \begin{axis}[
        scale=0.78,
        name=plot5,
        at={(plot4.east)},
        anchor=west,
        xshift=1cm,
        ybar,
        ymin=0,
        ymax=30,
        bar width=10pt,
        width=6.5cm,
        height=5cm,
        symbolic x coords={Critical, Ideal Goal, Rarely a Problem, Unknown},
        xtick=data,
        nodes near coords,
        nodes near coords align={vertical},
        enlarge x limits=0.15,
        ylabel={Number of Responses},
        xlabel={Importance of Reproducible Builds},
        tick label style={font=\tiny, rotate=45, anchor=east, align=center},
        title={(e) Reproducible Builds Importance},
        title style={font=\small\bfseries, yshift=5pt},
        every node near coord/.append style={font=\tiny},
    ]
    \addplot[fill=cyan!30] coordinates {(Critical,18)(Ideal Goal,27)(Rarely a Problem,9)(Unknown,4)};
    \end{axis}

    \begin{axis}[
        scale=0.78,
        name=plot6,
        at={(plot5.east)},
        anchor=west,
        xshift=1cm,
        ybar,
        ymin=0,
        ymax=25,
        bar width=10pt,
        width=6.5cm,
        height=5cm,
        symbolic x coords={Stability, Security, Balance, Not considered},
        xtick=data,
        nodes near coords,
        nodes near coords align={vertical},
        enlarge x limits=0.15,
        ylabel={Number of Responses},
        xlabel={Stability vs Security Tradeoff},
        tick label style={font=\tiny, rotate=45, anchor=east, align=center},
        title={(f) Stability vs Security Opinion},
        title style={font=\small\bfseries, yshift=5pt},
        every node near coord/.append style={font=\tiny},
    ]
    \addplot[fill=cyan!30] coordinates {(Stability,22)(Security,21)(Balance,11)(Not considered,4)};
    \end{axis}

    \end{tikzpicture}
    \caption{Survey Results on Dependency Management Practices and Challenges (n=58)}
    \label{fig:dependency-management-survey}
\end{figure}
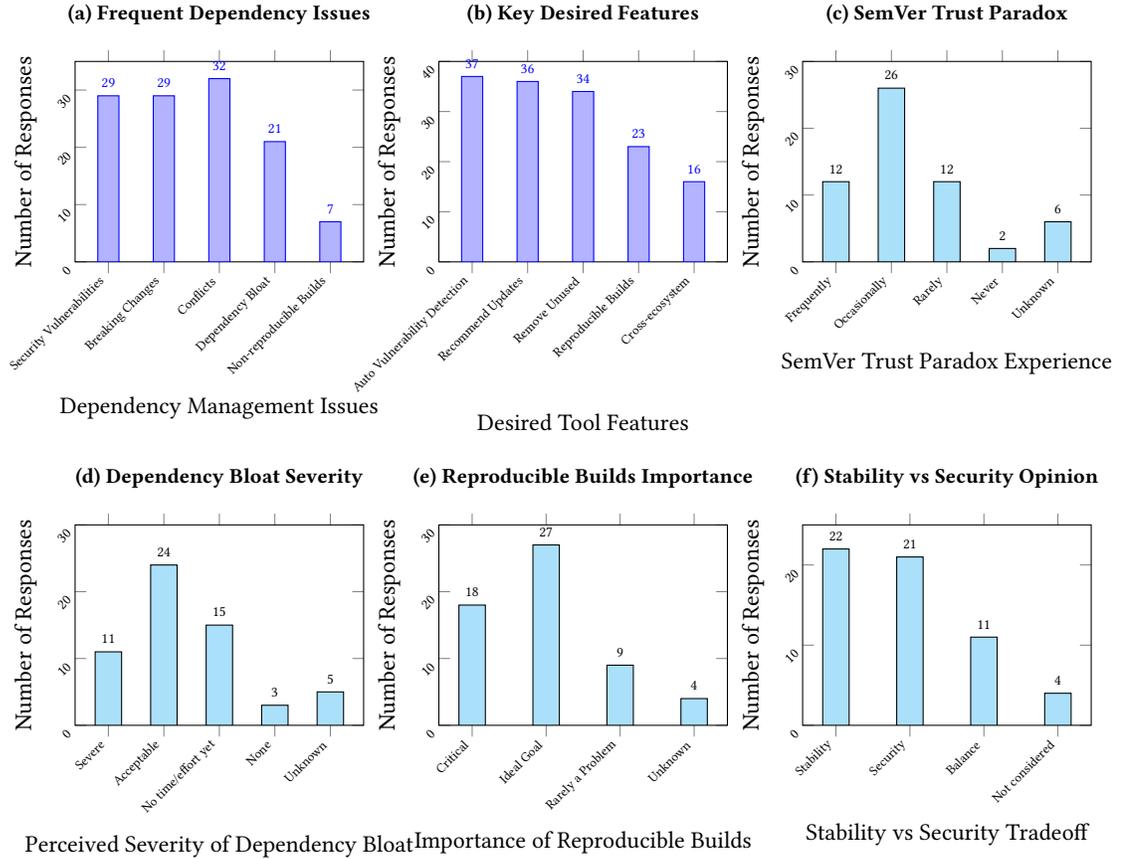

    The industrial feedback on this topic is presented in Fig.~\ref{fig:dependency-management-survey}. As with the previous topics, subfigures (a) and (b) in light purple concern the usage of automated tools, while the remaining subfigures in light blue present feedback on academic research areas.
    
    \subsubsection{Feedback on the Application of Automated Tools.} Our survey identified \emph{several critical challenges} in dependency management, ranked by the number of votes as follows (subfigure (a)): \textbf{\emph{dependency conflicts (55.17\%), vulnerability issues (e.g., known CVEs, 50\%), breaking changes (e.g., API incompatibility, 50\%), dependency bloat (introducing excessive, unused code, 36.21\%), and non-reproducible builds (which produce inconsistent outputs across different environments, 12.07\%).}} 
    About software dependency management tools (subfigure (b)), \emph{the most requested features} from participants include automated identification of \textbf{\emph{security vulnerability propagation paths (63.79\%), recommendations for compatible dependency updates (62.07\%), detection and removal of unused dependencies (58.62\%), ensuring reproducible builds (39.66\%), and cross-ecosystem dependency management (e.g., for C++/Rust, 27.59\%)}}. These challenges and feature request define the key problems that academic research should delve into.

   \subsubsection{Feedback on Academic Focus.} We gathered practitioner perspectives on four academic research topics.

First, we explored \emph{violations of Semantic Versioning (SemVer)}, where prior research indicates that approximately 28.6\% of non-major releases introduce breaking changes in the Golang Ecosystem~\cite{li2023large}. When asked about this ``trust paradox'', as shown in the subfigure (c), 44.83\% of participants encountered it occasionally, 20.69\% frequently, and 20.69\% rarely, \textbf{\emph{affirming the topic's practical relevance}}.

Second, regarding \emph{dependency bloat}, which reportedly affects over 50\% of dependencies on PyPI~\cite{drosos2024bloat}, only 18.97\% of participants considered it a serious issue requiring urgent attention (subfigure (d)). A larger share (41.38\%) acknowledged its presence but tolerated the impacts, while 25.86\% recognized the problem but lacked the resources to address it. Overall, \textbf{\emph{these responses suggest that despite its documented prevalence, dependency bloat is often deprioritized in practice.}}

Third, on the topic of \emph{reproducible builds}, research points to stark cross-ecosystem disparities, such as PyPI’s 12.2\% reproducibility success rate~\cite{benedetti2025empirical}. As shown in the subfigure (e), while most respondents viewed reproducible builds as an idealistic goal with prohibitive implementation costs, 15.52\% reported frequently encountering issues traceable to non-reproducibility. \textbf{\emph{This indicates a generally low perceived urgency among practitioners, despite tangible impacts experienced by a minority.}}

Finally, concerning the \emph{stability–security trade-off}, participant perspectives were divided (subfigure (f)): 37.93\% prioritized stability and avoided breaking changes, 36.21\% prioritized security and accepted associated risks, and 18.97\% struggled to balance both objectives. \textbf{\emph{This lack of consensus underscores the academic and practical value of further research into balancing these competing concerns.}}

    
    \vspace{3mm}
    \begin{custommdframed}
    \textbf{Findings on software dependency management in industry:}  Our study concludes that: 
    (1) The top three industrial challenges in dependency management are dependency conflicts, security vulnerabilities, and breaking changes. 
    (2) In response, the most critical features for automation are identifying security vulnerability propagation paths and facilitating compatible dependency updates.
    (3) Practitioners acknowledge the value of academic work on breaking changes and the stability-security balance, whereas topics like dependency bloat and reproducible builds receive less immediate interest.
    \end{custommdframed}
    \vspace{0cm}

    \subsection{Pre-trained Code Models (37 Responses).} 

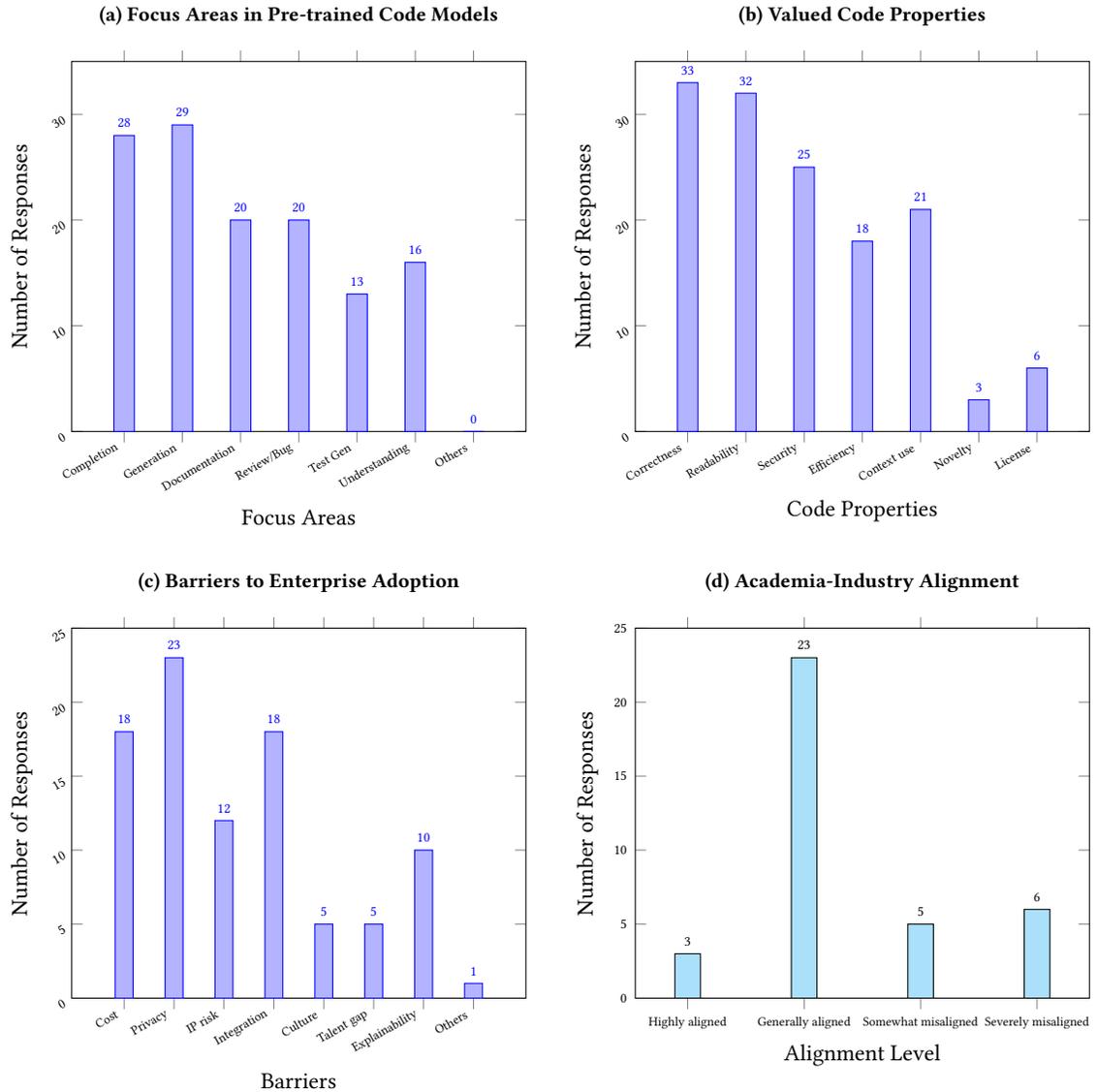
\begin{figure}[!htbp]
    \centering
    \begin{tikzpicture}
    
    \begin{axis}[
        scale=1.15,
        name=plot1,
        at={(0,0)},
        ybar,
        ymin=0,
        ymax=35,
        bar width=8pt,
        width=7cm,
        height=6cm,
        symbolic x coords={Completion, Generation, Documentation, Review/Bug, Test Gen, Understanding, Others},
        xtick=data,
        nodes near coords,
        nodes near coords align={vertical},
        enlarge x limits=0.15,
        ylabel={Number of Responses},
        xlabel={Focus Areas},
        tick label style={font=\tiny, rotate=30, anchor=east, align=center},
        title={(a) Focus Areas in Pre-trained Code Models},
        title style={font=\small\bfseries, yshift=5pt},
        every node near coord/.append style={font=\tiny},
    ]
    \addplot coordinates {(Completion,28)(Generation,29)(Documentation,20)(Review/Bug,20)(Test Gen,13)(Understanding,16)(Others,0)};
    \end{axis}

    \begin{axis}[
        scale=1.15,
        name=plot2,
        at={(plot1.east)},
        anchor=west,
        xshift=1.5cm,
        ybar,
        ymin=0,
        ymax=35,
        bar width=8pt,
        width=7cm,
        height=6cm,
        symbolic x coords={Correctness, Readability, Security, Efficiency, Context use, Novelty, License},
        xtick=data,
        nodes near coords,
        nodes near coords align={vertical},
        enlarge x limits=0.15,
        ylabel={Number of Responses},
        xlabel={Code Properties},
        tick label style={font=\tiny, rotate=30, anchor=east, align=center},
        title={(b) Valued Code Properties},
        title style={font=\small\bfseries, yshift=5pt},
        every node near coord/.append style={font=\tiny},
    ]
    \addplot coordinates {(Correctness,33)(Readability,32)(Security,25)(Efficiency,18)(Context use,21)(Novelty,3)(License,6)};
    \end{axis}

    \begin{axis}[
        scale=1.15,
        name=plot3,
        at={(plot1.south)},
        anchor=north,
        yshift=-2.7cm,
        ybar,
        ymin=0,
        ymax=25,
        bar width=7pt,
        width=7cm,
        height=6cm,
        symbolic x coords={Cost, Privacy, IP risk, Integration, Culture, Talent gap, Explainability, Others},
        xtick=data,
        nodes near coords,
        nodes near coords align={vertical},
        enlarge x limits=0.15,
        ylabel={Number of Responses},
        xlabel={Barriers},
        tick label style={font=\tiny, rotate=30, anchor=east, align=center},
        title={(c) Barriers to Enterprise Adoption},
        title style={font=\small\bfseries, yshift=5pt},
        every node near coord/.append style={font=\tiny},
    ]
    \addplot coordinates {(Cost,18)(Privacy,23)(IP risk,12)(Integration,18)(Culture,5)(Talent gap,5)(Explainability,10)(Others,1)};
    \end{axis}

    \begin{axis}[
        scale=1.15,
        name=plot4,
        at={(plot3.east)},
        anchor=west,
        xshift=1.5cm,
        ybar,
        ymin=0,
        ymax=25,
        bar width=10pt,
        width=7cm,
        height=6cm,
        symbolic x coords={Highly aligned, Generally aligned, Somewhat misaligned, Severely misaligned},
        xtick=data,
        nodes near coords,
        nodes near coords align={vertical},
        enlarge x limits=0.15,
        ylabel={Number of Responses},
        xlabel={Alignment Level},
        tick label style={font=\tiny, rotate=0},
        title={(d) Academia-Industry Alignment},
        title style={font=\small\bfseries, yshift=5pt},
        every node near coord/.append style={font=\tiny},
    ]
    \addplot[fill=cyan!30] coordinates {(Highly aligned,3)(Generally aligned,23)(Somewhat misaligned,5)(Severely misaligned,6)};
    \end{axis}

    \end{tikzpicture}
    \caption{Survey Results on Pre-trained Code Models Development and Adoption (n=37)}
    \label{fig:combined-code-models-survey}
\end{figure}

    The feedback on pre-trained code models is shown in Fig.~\ref{fig:combined-code-models-survey}. The first three subfigures pertain to the usage of automated tools, while the last one presents feedback on a specific academic focus.
    
    \subsubsection{Feedback on the Application of Automated Tools.}  The survey reveals the primary tasks targeted by automated tools among participants, namely code generation (78.38\%), code completion (75.68\%), documentation generation (54.05\%), code review and bug detection (54.05\%), code explanation (43.24\%), and test case generation (35.14\%) (Fig.~\ref{fig:combined-code-models-survey}(a)). \textbf{\emph{This task distribution shows strong consistency with prevailing academic research themes.}} The evaluation metrics employed further highlight a distinct industrial perspective. As shown in Fig.~\ref{fig:combined-code-models-survey} (b), although compilation success and logical accuracy (89.19\%) are the most critical criteria, significant attention is also paid to code readability and maintainability (86.49\%) and security (67.57\%). The consideration of contextual relevance (56.76\%) and, notably, licensing compliance (16.22\%) underscores that \textbf{\emph{practical industrial requirements encompass not only functional correctness but also long-term code quality and legal safeguards.}}

        In addition, participants reported several obstacles to adopting pre-trained code models (shown in Fig.~\ref{fig:combined-code-models-survey} (c)). These include data privacy and security concerns (62.16\%), cost and budget constraints (48.65\%), lack of integration with existing tools (48.65\%), uncertainty regarding intellectual property rights and legal risks (32.43\%), weak controllability and explainability (27.03\%), shortage of skilled talent (13.51\%), and resistance due to established development habits (13.51\%).  Collectively, these obstacles highlight that \textbf{\emph{the primary barriers to industrial adoption are often non-technical, revolving around security, cost, integration, and legal compliance, rather than mere performance on specific tasks.}}

        \subsubsection{Feedback on Academic Focus.}  Regarding current academic hotpots such as efficiency~\cite{liu2023empirical,wang2023one} , robustness~\cite{10.1145/3691620.3695072,liu2023contrabert}, and security~\cite{nie2024decoding,chen2024promise}, as shown in Fig.~\ref{fig:combined-code-models-survey} (d), 62.16\% of respondents believe these research directions generally align with industrial needs. However, \textbf{\emph{they emphasize that industry expects more out-of-the-box solutions rather than purely theoretical or methodological advances.}} 13.51\% perceive a certain divergence, noting that academic research tends to be overly cutting-edge, while \textbf{\emph{industry prioritizes solving fundamental issues like performance and cost}}. 16.22\% see a significant misalignment, pointing out that some research problems—such as pursuing extremely high accuracy on public benchmarks or detecting extremely rare types of vulnerabilities—are of limited importance in current practical applications.

    \vspace{3mm}
    \begin{custommdframed}
    \textbf{Findings on pre-trained code models in industry:}  
    (1) Industrial adoption is contingent not only on the functional accuracy of pre-trained code models but critically on a suite of non-technical factors encompassing data security, economic viability, systems integration, and legal assurance. 
    (2) Despite the pace of academic innovation, there is a clear industrial demand for solutions that are production-ready, with a emphasized prerequisite that fundamental issues of computational performance and cost are effectively addressed.
    \end{custommdframed}
    \vspace{0cm}

\section{Implications}
\label{sec:implications}

Through a comparative analysis of academic research and industrial needs, combined with feedback from the industry on academic research obtained from surveys, we have derived the following comparative findings for the further research.

\begin{itemize}[leftmargin=1em]
    \item \textbf{Significant Attention on Intelligent Software Requirements and Architecture Research are Required.} 
    \begin{itemize}[leftmargin=1.2em]
        \item \textbf{Evidence:} As illustrated in Fig.~\ref{fig:bar_chart}, research on software requirements and architecture remains the least prevalent in top software engineering conferences. This scarcity does not indicate that automated techniques for these two phases are mature. On the contrary, requirements engineering and architectural design still largely depend on manual expert effort ~\cite{DBLP:journals/pacmse/LianWZLWZ25, DBLP:conf/re/LianMLZ24, 10971683,bucaioni2025artificialintelligencesoftwarearchitecture}. This research gap is primarily due to challenges such as low transparency of public benchmarks, high diversity in representations, and a strong dependence on tacit domain-specific knowledge and experience ~\cite{chen2025deep}.
        \item \textbf{Action for researchers:} The importance of software requirements and architecture for shaping high-quality software is undoubted. They ensure that software development ``does the right thing'' and ``does the thing right'' from an overarching perspective, respectively. This is especially critical with the rapid development of intelligent code synthesis and testing, which now require higher-quality requirements and more intelligent architectural design than before, as the opportunity for validation by domain experts with tacit knowledge is reduced. Consequently, \emph{advanced techniques are needed to automatically synthesize sound requirements, detect and repair requirement smells (e.g., conflicts, incompleteness, ambiguity), and recommend architectural designs that decompose systems into modules implementable by current code LLMs.}
    \end{itemize}

    \item \textbf{Reliability and explainability are paramount for the successful adoption of LLM-augmented approaches in software engineering.} 
    \begin{itemize}[leftmargin=1.2em]
        \item  \textbf{Evidence:} For example, in the program analysis task (Section~\ref{subsec: programAnalysis}), only 7.5\% of participants expressed high confidence in using LLMs to augment traditional program analysis. The remaining respondents doubted their effectiveness due to hallucinations, despite acknowledging their potential. They emphasized the need for stricter verification of LLM outputs using conventional methods to mitigate this issue. Additionally, practitioners seek more explanatory information about the decisions made by LLMs, enabling better human evaluation of their trustworthiness (Section~\ref{subsec:programRepair}).

        \item \textbf{Action for researchers:}  In response, researchers have made initial progress in mitigating hallucinations. For example, Liu et al. established a taxonomy for hallucinations in functional code ~\cite{liu2024exploringevaluatinghallucinationsllmpowered}, while Zhang et al. investigated their phenomena and mitigation in repository-level code generation ~\cite{zhang2025llmhallucinationspracticalcode}. Additionally, Liu et al. empirically demonstrated the robustness weaknesses of state-of-the-art program generation models ~\cite{10.1145/3641540}. However, this research is currently concentrated on code synthesis tasks, leaving areas like requirements analysis, maintenance, and testing largely unaddressed. \emph{These studies also remain preliminary, underscoring the urgent need for more advanced techniques to ensure reliability.}

        Simultaneously, a growing body of research aims to make LLM decisions more interpretable by providing explanations ~\cite{10.1145/3691620.3695071, 10.1145/3611643.3616309, 10.1145/3660814, 10.1145/3660771, 10.1109/ICSE48619.2023.00063, DBLP:conf/icse/Cao0W0B0024}, such as the rationale or similar historical cases. \emph{According to our findings, practitioners value this transparency as it helps them decide whether to trust an automated output.} We encourage more work in this direction to enhance the practical adoption and positive social impact of LLMs in software engineering.
    \end{itemize}

    \item \textbf{Divergent Inputs for Test Case Generation in Academic and Industrial.} 
    \begin{itemize}[leftmargin=1.2em]
        \item \textbf{Evidence:} According to industry surveys, software requirements specifications remain the most common input for creating test cases (70.27\%). However, in academic research, test generation predominantly relies on source code and testing requirements, creating a notable divergence between industry practice and research focus~\cite{rao2023cat,yin2025you,lin2023automated,sun2023revisiting}. This gap threatens the practical relevance and applicability of academic outputs in real-world settings.

        Effective testing requires software requirements that are correct, complete, and consistent. Yet, achieving these qualities remains highly challenging in practice~\cite{skokovic2010requirements,mustafa2021automated}. Several factors contribute to this difficulty: the cognitive limitations of requirements engineers~\cite{DBLP:conf/re/ZhaoZL21}, the inherent ambiguity of natural language in documenting requirements~\cite{9793957}, and the often tacit knowledge held by domain experts~\cite{DBLP:journals/spe/ZhaoZL24}.
    
     During our survey, five testing experts reported a common pain point:
    ``\emph{We are tasked with deriving test requirements from software requirements, yet we frequently encounter severe issues—such as incompleteness, inconsistencies, and ambiguities. When we report these issues to the requirements engineers, the subsequent revisions force adjustments across both development and testing activities. This results in a prolonged and tedious cycle. There is a strong need for automated techniques and tools capable of identifying and rectifying weaknesses in requirements specifications.}''

        \item \textbf{Action for researchers:}  Two urgent industrial needs emerge: (1) software requirements quality assurance technologies and tools, and (2) software requirements specification (SRS)-based testing technologies.

        For requirements quality assurance, while there is initial research exploring requirement smell classification~\cite{alem2025multi}, smell detection~\cite{FEMMER2017190}, and missing requirements recommendation~\cite{DBLP:journals/pacmse/LianWZLWZ25}, existing approaches remain limited. They are not yet systematic, often fail to alert practitioners to multiple co-occurring smells, and lack the ability to automatically generate high-quality modifications. This underdevelopment is also evidenced by our academic analysis (Section~\ref{sec:academicAnalysis}) and corroborated by existing literature reviews~\cite{chen2025deep}. \emph{To address this, we call for greater academic attention toward developing practical, industrial-grade requirements quality assurance technologies and tools that can systematically detect, analyze, and rectify specification flaws at scale.}

        Turning to SRS-based testing, the field has historically depended on a precise, machine-readable interpretation of requirements, leading to work on formal~\cite{10.5555/857174.857305} and model-based~\cite{9491761, mohdshafie2022model} specifications. While a small body of work attempts to generate tests directly from natural language SRS, it struggles with producing highly relevant test suites and ensuring the inclusion of critical test cases ~\cite{DBLP:journals/corr/abs-2412-03693}. This is largely due to the profound challenge of automated understanding nuanced system behaviors and broader project context from unstructured text. Consequently, we advocate for research that bridges this gap, particularly through hybrid approaches that combine natural language processing with structured models (e.g., UML) for deeper requirements understanding and more robust, context-aware test generation.
    \end{itemize}

    \item \textbf{The proliferation of multi-language software systems in large-scale projects has created a pressing demand for cross-language program analysis.} 
    \begin{itemize}[leftmargin=1.2em]
        \item \textbf{Evidence:}  Contemporary development frequently leverages the strengths of various languages, for instance, employing C/C++ for performance-critical components and Python for high-level control and integration. This paradigm, though flexible, introduces substantial complexity into debugging and program analysis. In response to these challenges, four significant works have emerged in top conference forums over the past three years. These studies have pioneered static analysis for JavaScript-Java communication ~\cite{10298294} ~\cite{10.1145/3691620.3696193}, performed pointer analysis across Java and C ~\cite{11029730}, and addressed undefined behavior in Rust libraries by examining the intricacies of its ownership model at language boundaries (e.g., with C/C++) ~\cite{11029832}. 

        \item \textbf{Action for researchers:} It is obvious that the study on this topic just starts, constrained by their narrow focus on specific language pairs and a limited set of communication mechanisms. This resulting in fragmented solutions that struggle to generalize across the diverse and evolving landscape of polyglot systems. To address these limitations, \emph{future research should expand its scope to encompass a wider variety of programming languages, such as Go, WebAssembly, and Python-native extensions, and broader interoperability paradigms, including modern RPC frameworks and memory-sharing models.}
    \end{itemize}

    \item \textbf{The automatic synthesis of fundamental software systems, such as operating system kernels and drivers, remains a largely unexplored frontier.} 
    \begin{itemize}[leftmargin=1.2em]
        \item \textbf{Evidence:}  While existing benchmarks have driven progress in code synthesis, they are predominantly confined to stand-alone functions involving basic programming tasks (e.g., HumanEval) or competitive programming problems (e.g., APPS, CodeContest). Although a few repository-level datasets exist (e.g., RepoEval, CoderEval), they focus primarily on high-level applications written in languages like Java or Python. Consequently, the automatic synthesis of critical low-level components, such as OS drivers which are in constant industrial demand due to frequent hardware changes, represents a critical and unaddressed challenge.

        \item \textbf{Action for researchers:} We urge researchers to pursue the synthesis of high-assurance, requirement-intensive code, particularly in domains such as operating system components (e.g., drivers) or industrial control systems (e.g., PLC programs). These systems impose stringent, multi-faceted requirements that extend beyond functional correctness, including safety, security, and real-time constraints. \emph{As an initial step, we call for the creation of domain-specific, industrially relevant benchmarks that capture these rich requirements. These benchmarks should then drive the development of novel synthesis methods and evaluation frameworks capable of meeting the rigorous standards of safety-critical software engineering.}
         
    \end{itemize}

    \item \textbf{Rethinking the Evaluation of Code Generation and Completion for Practical Impact.}
    \begin{itemize}[leftmargin=1.2em]
        \item \textbf{Evidence:} A critical gap persists between academic research and industrial application in code generation and completion, largely due to a misalignment in priorities. While academia actively tackles complex challenges, such as project-level context understanding, code quality, and security, using advanced architectures like RAG and multi-agent systems, its evaluation paradigm remains confined to controlled benchmarks (e.g., JavaBench ~\cite{cao2024javabench}, RepoEval ~\cite{zhang2023repocoder}) and narrow metrics such as pass@k, BLEU and Exact Match~\cite{yu2024codereval,wang2024rlcoder}. In contrast, \emph{industrial adoption demands out-of-the-box robustness, seamless integration into development environments, and cost efficiency, which are often overlooked in research settings.}

        This disconnect manifests in practical shortcomings: for instance, generated class-level code frequently requires extensive manual modification, and models often fail to recognize deprecated APIs. Beyond pass rates, industry prioritizes inference speed, workflow compatibility, and long-term maintainability. Thus, the core challenge lies in transitioning from isolated research breakthroughs to systematic, cost-effective, and production-ready solutions.
        
        \item \textbf{Action for researchers:}  Closing this gap will require academia to adopt more comprehensive and practical evaluation frameworks, incorporating metrics like adoption rate, robustness, and explainability, and to demonstrate real-world utility through seamless integration with mainstream IDEs, thereby amplifying the social impact of research.
    \end{itemize}

    \item \textbf{Ethical Considerations: An Emerging Requirement for Pre-Trained Code Models}
    \begin{itemize}[leftmargin=1.2em]
        \item \textbf{Evidence:} Current research on pre-trained code models predominantly focuses on enhancing efficiency and effectiveness, addressing areas such as code representation and understanding ~\cite{10298296,10.1145/3691620.3695072,10.1145/3691620.3694999,10172499,10172503}, long-tailed code distribution ~\cite{10298393}, memorization ~\cite{10.1145/3729390, 10.1145/3597503.3639074} and model lightweight ~\cite{10.1145/3729405}. However, our survey indicates a significant shift in practitioner priorities toward pressing ethical concerns. \emph{These include the protection of private data, intellectual property risks, and legal liabilities arising from generated code. These issues may lead to more severe consequences than the incremental gains in human productivity. }
                       
        \item \textbf{Action for researchers:}  Although preliminary studies, such as Zhou et al.'s investigation into memorization in LLMs ~\cite{10.1145/3597503.3639074}, have begun to explore related aspects, the field remains in its infancy. \emph{There is a clear need for more robust mechanisms, strategies, and techniques to prevent training data leakage and mitigate associated risks.}

        Furthermore, the overconfidence of code models presents a significant practical challenge, as they typically generate answers even when operating outside their valid knowledge domain. This behavior necessitates that practitioners constantly remain vigilant, without clear guidance on which outputs require heightened scrutiny. To address this issue, \emph{there is a strong demand for models capable of estimating and explicitly conveying their uncertainty alongside each generated response towards specific SE tasks.} Providing such confidence indicators would enable developers to quickly identify and critically evaluate potentially unreliable code suggestions, thereby integrating AI assistance into workflows more safely and efficiently.
    \end{itemize}

\end{itemize}

\section{Discussion}
\label{sec:discussion}

\subsection{Threats to Validity}
\emph{Internal Validity} concerns factors that could influence the causal interpretation of our results. A primary threat was the potential for insincere responses to the questionnaire. To mitigate this, we discarded all responses completed in less than one minute, a threshold established through a pilot study to safeguard data quality.

\noindent \emph{Construct Validity} refers to how well our measurements capture the intended theoretical concepts. A threat stems from our questionnaire design, which was structured around specific themes to address the research goal and might have constrained broader feedback. We mitigated this by designing the questions based on well-established hotspots and challenges from the literature, ensuring relevance, and keeping the survey within a tolerable time frame to maintain participant engagement. A further threat is that our operationalization of ``academic capability'' relies on performance from public benchmarks, which may not reflect real-world utility.

\noindent \emph{External Validity} concerns the generalizability of our findings. A threat to this is our publication selection from the top-three conferences, excluding journals and other forums. We contend that the 1,367 publications from these premier venues, which attract the best research, provide a representative snapshot of the community's advanced techniques. Similarly, the industrial feedback is from a specific set of participants and may not represent the entire industry. To mitigate this threat, we solicited responses from a diverse range of companies (e.g., Microsoft, ByteDance, Momenta) and developer roles (e.g., software engineers, algorithm engineers, product managers).

\subsection{Limitations}

\begin{itemize}[leftmargin=1em]
    \item \textbf{Potential Limitations in Assessing Model Capabilities.} A key limitation of our methodology is its dependence on performance metrics reported in the original studies, which are confined to the common benchmarks prevalent in the selected SE conferences. While this allows for a consistent comparison across different approaches, it presents two challenges. The reproducibility of these results is uncertain, often due to insufficient details on experimental parameters and the inherent randomness of LLMs. Furthermore, this focus may cause us to miss specialized capabilities that are only visible on novel, customized benchmarks, thereby narrowing the scope of our analysis.

    \item \textbf{Potential Limitations in Industrial Survey.} First, despite our efforts to recruit practitioners from 17 diverse groups across various software domains, our reliance on social networks for sampling (a non-probability method) and the predominance of respondents from China may limit the geographical and cultural generalizability of our findings. Second, while the approximately 40 responses per topic came from mature software engineers, whom we believe to be a representative cohort, the modest sample size means that a larger study could yield more nuanced and robust conclusions.
\end{itemize}

\section{Related Work}

Recognizing the importance of aligning industrial needs with academic research, scholars have conducted empirical studies to examine the barriers to adopting intelligent techniques for specific tasks from multiple perspectives.

Stradowski and Madeyski ~\cite{stradowski2023bridging} analyzed Nokia's experience with ML-based defect prediction. They identified key contextual factors influencing adoption and distilled their findings into thirteen holistic considerations. Laiq et al. ~\cite{laiq2024industrial} conducted an industrial case study of an ML-based bug triage tool (invalid bug report prediction) and identified key adoption challenges including: practitioners demanded interpretable predictions, seamless integration with existing toolchains, and ongoing model maintenance (retraining) to sustain accuracy over time. Likewise, Syu and Wang ~\cite{syu2023gap} report a disconnect in automated service composition (ASC): academic ASC tools assume structured (tuple-based) service descriptions, whereas industry engineers use free natural-language descriptions. This mismatch – requiring costly manual translation – blocks adoption.  Wang et al. ~\cite{wang2023practitioners} conducted a large-scale survey of 599 developers from 18 IT companies to investigate practitioner expectations on code completion, including usage scenarios, evaluation metrics, and key adoption factors such as efficiency and effectiveness; Li et al. ~\cite{li2025understanding} conducted a qualitative study to evaluate static application security testing (SAST) tools through in-depth, semi-structured interviews with 20 industrial professionals, aiming to understand their motivations, challenges, and expectations regarding SAST benchmarks and tool evaluation practices; Steenhock et al. ~\cite{steenhoek2024closing} conducted a user study with 17 professional developers from Microsoft across 24 projects to assess the usability of an \emph{AI-based vulnerability detection and repair} tool integrated into IDEs.  Yu et al. ~\cite{yu2024practitioners} interviewed 13 professionals and surveyed 339 practitioners on automated test generation, and found that decades of research in test generation have not met developers’ expectations, and they outline new directions to better align research tools with industry needs. 

Beyond these domain-specific investigations, broader analyses of the research–practice gap have been conducted. Lo et al. ~\cite{lo2015practitioners} conducted a large-scale survey of 512 Microsoft engineers to assess the perceived relevance of software engineering research across 571 FSE and ICSE papers in 2015, revealing that while 71\% of research ideas were rated as essential or worthwhile, significant gaps remained in aligning academic work with practitioners’ needs. In a systematic review, Brings et al. ~\cite{brings2018approaches} catalogued technology-transfer approaches in SE and found that successful adoption usually involves direct academia–industry collaboration. Crucially, they note that technological maturity and adaptability of a solution are necessary preconditions for transfer, whereas social and organizational mismatches are major barriers. Similarly, Garousi et al. ~\cite{garousi2020practical} synthesized decades of community feedback on research relevance and concluded that simplistic academic assumptions and misaligned problem selection are root causes of the gap. They recommend more industry-collaborative, problem-driven research approaches to increase impact. Besides, some efforts have aimed to narrow the gap between academic research and industrial practice. For example, ResearchBot \cite{11050792} was developed to support programming communities by automatically retrieving and synthesizing information from relevant academic publications.

Building on these efforts to bridge research and practice, recent studies have also begun to evaluate how emerging technologies such as large language models (LLMs) are assessed within software engineering. Hu's research ~\cite{hu2025assessing} examines 291 benchmarks developed for assessing LLM performance across diverse SE tasks, including requirements engineering, coding assistance, software testing, AIOps, software maintenance, and quality management. This benchmark research not only catalogs available evaluation resources but also critically analyzes their construction methodologies, highlights limitations of existing benchmarks, and discusses future challenges and opportunities for SE-related evaluation tools, thereby providing a foundational overview for developing more effective assessment frameworks in this rapidly evolving field. Besides, Jiang et al.~\cite{jiang2024survey} conducted a comprehensive survey on code-oriented LLMs (Code LLMs), providing the first systematic literature review dedicated to LLMs for code generation. Their study proposes a taxonomy encompassing key aspects such as data curation, model advances, performance evaluation, ethical considerations, and real-world applications. By aggregating both quantitative and qualitative comparisons across widely used benchmarks (e.g., HumanEval, MBPP, and BigCodeBench~\cite{zhuo2024bigcodebench}), the survey underscores the continuous advancement of LLM capabilities in code generation while pinpointing key challenges and emerging opportunities to narrow the gap between academic research and practical development.

While existing studies offer valuable domain- or company-specific insights, they often focus on isolated tasks (e.g., defect prediction ~\cite{stradowski2023bridging}, service composition ~\cite{syu2023gap}, code completion ~\cite{wang2023practitioners}) or individual tools (e.g., static application security testing ~\cite{li2025understanding}, AI-based vulnerability detection and repair ~\cite{steenhoek2024closing}). Consequently, they lack a holistic view of the broader academia-industry relationship in software engineering. Although our prior study ~\cite{lo2015practitioners} provided a comprehensive analysis, it is now a decade old. To refresh these findings, this work conducts a large-scale, data-driven, cross-domain investigation that quantitatively links academic research trends with industrial needs across six major SE domains. By integrating bibliometric analysis of top-tier publications with targeted industrial surveys, we propose a unified empirical framework. This framework complements perception-based studies like Lo et al. ~\cite{lo2015practitioners} with systematic, topic-level evidence spanning diverse intelligent techniques and their benchmark performance. In doing so, it extends the research of Hu et al. ~\cite{hu2025assessing} and Jiang et al. ~\cite{jiang2024survey}, offering a holistic understanding of academia-industry alignment and providing actionable insights for future SE research.
\label{sec:relatedWork}

\section{Conclusion}
\label{sec:conclusion}

Academic research serves as a primary driver of industrial progress. Periodically aligning academic pursuits with industrial needs is crucial to ensure that \emph{academia is tackling real-world problems} and to \emph{redirect research attention toward significant yet under-explored areas}. To this end, this study presents a systematic investigation into academic capabilities, industrial feedback, and future research directions. We collected and analyzed 1,367 publications from three premier SE conferences (ASE, FSE and ICSE), mapping their distribution across software development phases and identifying prominent research hotspots. Through an in-depth analysis of six major topics anchored by common benchmarks, we derived 11 key findings. We further conducted topic-specific surveys, gathering 282 responses from 17 organizations, which revealed current challenges in adopting automated techniques and outlined industrial expectations from a practical standpoint. Based on the interplay between academic advances and industrial insights, we propose seven key implications, aiming to inform and guide the future trajectory of software engineering research.

\section*{Data Availability}
The annotated sheets of the academic papers and the corresponding questionnaire feedback on six hot academic topics are provided at \url{https://github.com/yuhangbuaa/Aligning-Academia-with-Industry.git}.

\section*{Acknowledgments}
Funding for this work was provided by the National Natural Science Foundation of China (Grants No. 62177003).

\bibliographystyle{ACM-Reference-Format}
\bibliography{Reference}
\end{document}